\documentclass[twocolumn,showpacs,preprintnumbers,amsmath,amssymb,floatfix]{revtex4}

\usepackage{psfig}
\usepackage{epsf}
\usepackage{graphicx}% Include figure files
\usepackage{dcolumn}% Align table columns on decimal point
\usepackage{bm}% bold math

\begin{document}

\title{CO adsorption on close-packed transition and noble metal surfaces: Trends from ab-initio calculations}
% Force line breaks with \\

\author{Marek Gajdo\v s}
 \email{Marek.Gajdos@univie.ac.at}
\author{Andreas Eichler}%
\author{J\"urgen Hafner}%
\affiliation{%
Institut f\"ur Materialphysik and
Center for Computational Materials Science\\
Universit\"at Wien, Sensengasse 8/12, A-1090 Wien, Austria
}%

\date{\today}% It is always \today, today,

\begin{abstract}
We have studied the trends in CO adsorption on close-packed metal
surfaces: Co, Ni, Cu from the 3d
row, Ru, Rh, Pd, Ag from the 4d row and Ir, Pt, Au from the 5d row using density functional theory.
In particular, we were concerned with the trends in the adsorption energy, the geometry, the vibrational properties and other parameters derived from the electronic structure of the substrate.  The influence of specific changes in our setup such as choice of the
exchange correlation functional, the choice of pseudopotential and size of the basis set, substrate relaxation has been carefully evaluated. We found that while the geometrical and vibrational properties of the adsorbate-substrate complex are calculated with high accuracy, the adsorption energies calculated with the gradient-corrected Perdew-Wang exchange-correlation energies are overestimated. In addition, the calculations tend to favour adsorption sites with higher coordination, resulting in the prediction of wrong adsorption sites for the Rh, Pt and Cu surfaces (hollow instead of top). The revised Perdew-Burke-Erzernhof functional (RPBE) leads to lower (i.e. more realistic) adsorption energies for transition metals, but to wrong results for noble metals - for Ag and Au endothermic adsorption is predicted. The site preference remains the same. We discuss trends in relation to the electronic structure of the substrate across the Periodic Table, summarizing the state-of-the-art of CO adsorption on close-packed metal surfaces.
\end{abstract}

%\keywords{Suggested keywords}%Use show keys class option if keyword
\maketitle

\vspace{0.2cm}
\section{Introduction}
\label{introduction} The development of modern theoretical
surface science provides an opportunity to investigate surfaces and
adsorbate structures on the atomic scale with useful applications
in industrial technologies. Much effort has been devoted to
study CO chemisorption and dissociation on transition metals.
There are numerous papers and reviews which deal with this system
from different points of view (electronic, structural,
vibrational) \cite{jpc:Blyholder:68,prb:Bagus:28,pss:Over:58,jacs:Sung:107}.
One of the central questions is how the strength of the
chemisorption of CO and the preference for the specific adsorption
site varies across the transition metal (TM) series. A particular
case, CO adsorption on the Pt(111) surface, has attracted much
attention in the past, since for this system state-of-the-art DFT
calculations fail in predicting the correct site preference
\cite{jpcb:Feibelman:105}. The question which immediately arises,
whether this case is an exception or the rule ?

Additionally, a vast number of theoretical papers appeared in the literature since Ying, Smith and Kohn
presented in the mid 70's a first self-consistent density functional study of chemisorption on metal surfaces (H on tungsten) \cite{prb:Ying:11}.
The number of theoretical adsorption studies of TM molecules on different surfaces is increasing with time not only because of their importance in catalysis, but also due to the increasing reliability of the ``measured" properties.
In the past, several systematic studies of the adsorption of CO molecules on transition metal surfaces have been performed \cite{prb:Bagus:28,prb:Andreoni:23,ac:Hammer:45,rpp:Norskov:53}.
In 1990 N{\o }rskov  proposed a model of chemisorption on transition metal surfaces which was later expanded and is now quite generally accepted \cite{rpp:Norskov:53,ac:Hammer:45}.  A main feature of the model is the importance of the position of the d-band center relative to the HOMO and LUMO of the adsorbate.
The importance of understanding the correlation between the geometric and the electronic structure arises from the proposed mechanistic  model for chemisorption.
One among several trends is the correlation between the CO stretching frequency $\nu_{\rm C-O}$ and the energy level difference between $5\sigma$ and $1\pi$ orbitals of the adsorbed CO ($\Delta(5\tilde{\sigma}-1\tilde{\pi})$) proposed by Ishi et al. \cite{ss:Ishi:161}.

In this paper we present an extensive density functional study
of the adsorption of CO on the closed-packed surfaces [(111) for face-centered cubic, resp. (0001) for hexagonal metals] of Co, Ni, Cu, Ru, Rh, Pd, Ag, Ir, Pt and Au.
After a description of the set-up in Section \ref{Methodology}, we
characterize briefly the clean surfaces in Section \ref{Clean
surfaces}. Section \ref{CO adsorption} is devoted to the
adsorption of CO. Starting from the geometric structure of the
adsorbate-substrate system, going through vibrational and electronic
properties (stretching frequencies of the adsorbate, and of the adsorbate-substrate bond, occupation of the anti-bonding
2$\pi^{\star}$-like orbital, density of states, redistribution of the charge density) as well as site preference we draw a
complete  picture of the CO adsorption on the close-packed TM
surfaces. Moreover, we investigate the influence of the
exchange-correlation functional and the cut-off energy on the site preference (Section
\ref{Improving Site Preference}). In Section \ref{Discussion} we discuss our results in the light of
the experimental literature and analyze trends and correlations
between the investigated properties.
This paper tries to go beyond a pure table of references from
experiments and different DFT calculations, by providing a large and
consistent database, in which each element was treated in exactly
the same manner. Further, one of our main goals is not only to
obtain theoretical values of spectroscopic accuracy, but also to
derive useful trends of CO adsorption on 3d, 4d and 5d transition
metal surfaces with the hope of applicability in the prediction of
the adsorption and catalytic behavior.

\section{Methodology}
\label{Methodology}

The calculations in this work are performed using the Vienna
ab-initio simulation package VASP \cite{vasp,prb:Kresse:54} which
is a DFT code, working in a plane-wave basis set. The electron-ion
interaction is described using the projector-augmented-wave (PAW)
method \cite{prb:Blochl:50,prb:Kresse:59} with plane waves up to
an energy of E$_{\rm cut}$~=~450~eV (for some calculations harder
pseudopotentials were used for C and O which require energy
cut-off of 700 eV). For exchange and correlation the functional
proposed by Perdew and Zunger \cite{prb:Perdew:23} is used, adding
(semi-local) generalized gradient corrections (GGA) of various
flavor (PW91 \cite{prb:Perdew:46}, RPBE \cite{prb:Hammer:59}).
These GGAs represent a great improvement over the local density approximation (LDA) in the
description of the adsorption process.

\begin{figure}[htb]
\centerline{\psfig{figure=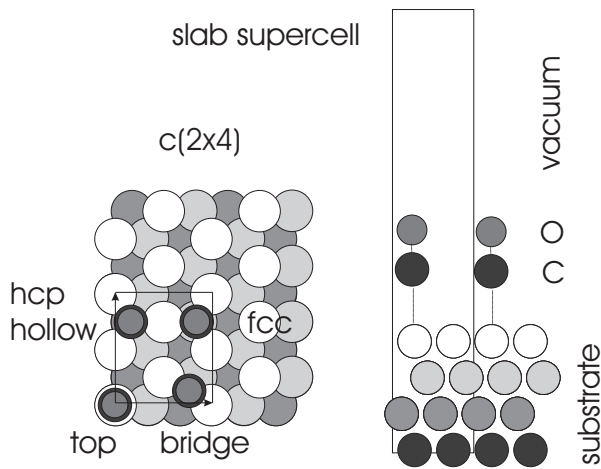,width=8.5cm,clip=true}}
\nopagebreak \caption{\label{figure structure} Top and side view of the slab used in the calculations. In the top view of the c(2$\times$4) cell we denote all investigated high-symmetry sites.}
\end{figure}

The substrate is modelled by four layers of metal separated by
a vacuum layer of approximately double thickness, as shown in
Fig. \ref{figure structure}. The two uppermost substrate layers
and the CO molecule are allowed to relax. This enables us to check
the influence of the relaxation on the adsorption system. The
Brillouin zone of the  c(2$\times$4) surface cell (equivalent to a
coverage of $\Theta$~=~0.25~ML) was sampled by a grid of
(4$\times$3$\times$1) k-points. We have chosen this coverage as a
compromise between small adsorbate-adsorbate interactions (``low
coverage limit") and low computational effort.

In the calculation we investigated the adsorption on the close
packed surfaces of 10 metallic elements from the 3d (Co, Ni, Cu),
4d (Ru, Rh, Pd, Ag) and 5d transition metal rows (Ir, Pt, Au) of
the Periodic Table. For all hcp-elements, Co
and Ru, the (0001) surface was used with the ideal $c/a$ ratio of 1.63. The spin-polarization of
Ni and Co was also taken into account.

Vibrational properties of CO were computed by applying a finite-differences
method to create the Hessian matrix which we
diagonalize to obtain the characteristic frequencies. We have
calculated the metal-CO ($\nu_{\rm M-CO}$) and C-O stretch ($\nu_{\rm C-O}$)
frequencies in the direction perpendicular to the surface plane.

The free CO molecule is characterized by a calculated stretch
frequency of 2136~cm$^{-1}$ at an equilibrium bond length of 1.142~\AA.
The corresponding experimental values are 2145~cm$^{-1}$ and 1.128~\AA
\cite{jms:Mantz:39}. The problem of a too large CO binding
energy (E$_{\rm CO}^{PW91}$~=~11.76~eV, E$_{\rm CO}^{exp}$~=~11.45~eV  \cite{book:Cox:1984}) stems
mainly from the error in the energy of the free atom, where high
density gradients make an accurate description more difficult.

Further, the bonding of adsorbate to the surface by calculating density of states and charge density flow was investigated.
\section{Bulk and Clean surfaces}
\label{Clean surfaces}

\subsection{Lattice constant}
\label{Lattice constant} 
\begin{figure}[htb]
\centerline{\psfig{figure=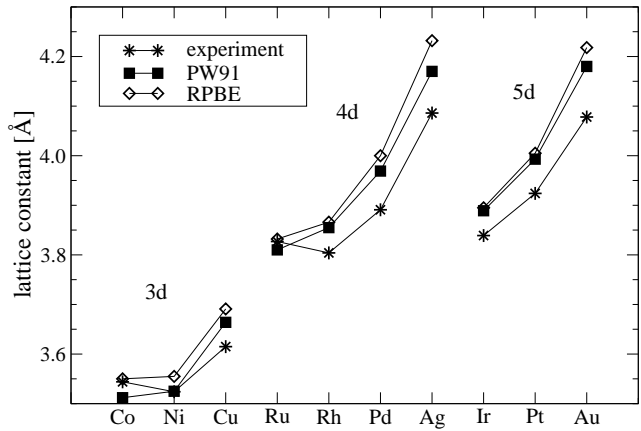,width=9.0cm,clip=true,angle=-90}}
\caption{\label{figure lattice} Lattice parameters for a part of the
Periodic Table. We show the experimental and calculated lattice parameters for PW91 and RPBE functional. Although, Ru and Co are hcp metals we have included them together with fcc metals at an ideal c/a~$\doteq$~1.63.}
\end{figure}

For a general understanding of the
adsorption process, especially for the adsorption in higher
coordinated sites, the lattice constant of the substrate is an important parameter.
 The optimal adsorption height for example will always be
 determined by an interplay between the optimal carbon-metal
 bond-length and the lattice parameter. For that reason, we give
 in Fig. \ref{figure lattice} the theoretical lattice constants
 together with the experimental values
\cite{springer:Landolt:1971}. These lattice parameters are compared
with lattice parameters calculated with the PW91 and RPBE exchange
correlation functional.

The overestimation of the lattice parameter (calculated within the
GGA) is characteristic for the heavier elements. The difference
between the experimental and calculated lattice constant is around
1\% for 3-d row metals and around 2\% for other elements. Such a  2\%
difference in the lattice constant corresponds to a change in the
CO adsorption energy of about 0.03~eV for Pd(100)
\cite{jcp:Wu:113} and Ru(0001) surface \cite{prl:Mavrikakis:81}. In general, the lattice parameter
increases along the rows and columns of the Periodic Table. For a
greater lattice constant one might expect that the adsorbate will
come closer to the surface. On the other hand, as the width of the d-band increases, the binding is reduced and the actual height of
the adsorbate over the surface should increase.

\subsection{Work-function}
\label{Workfunction}

\begin{figure}[htb]
\centerline{\psfig{figure=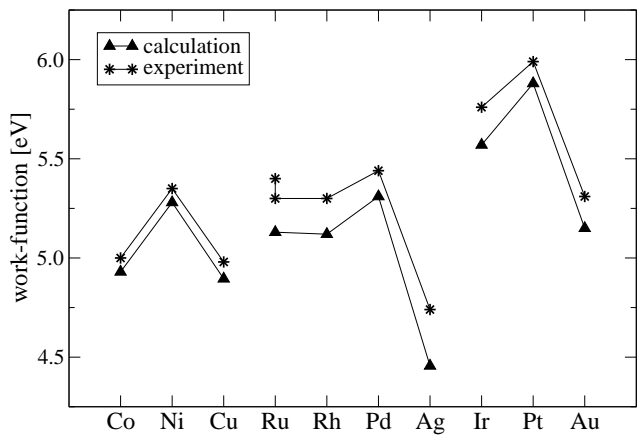,width=9.0cm,clip=true,angle=-90}}
\nopagebreak
\caption{\label{workfunction clean} Experimental and calculated values of the work-function (PW91) for various transition metal (111) surfaces. References for experiments: Co \cite{jap:Michaelson:48}, Ni \cite{jap:Michaelson:48}, Cu \cite{jap:Michaelson:48}, Ru \cite{ss:Himpsel:115,prb:Bottcher:60}, Rh \cite{phd:Hendrickx}, Pd \cite{prl:Fischer:70}, Ag \cite{jap:Michaelson:48}, Ir \cite{jap:Michaelson:48}, Pt \cite{ss:Nieuwenhuys:34,phd:Hendrickx}, Au \cite{jap:Michaelson:48}.}
\end{figure}

In Fig. \ref{workfunction clean} we present the calculated
work-functions together with experimental values.

We find that the calculated work-functions (PW91) are in all cases slightly lower
(by at most $6\%$ for Ag) than the measured values. The largest
discrepancies in terms of absolute values are found for 4d metals,
where the differences for Ru, Rh and Ag surface are
$\approx$0.3~eV. However, in general the agreement is quite good.
The work-function always increases with the d-band filling.
Only for the noble metals (Cu, Ag, Au) it decreases again.
Similarly, the work-function increases along the columns when
going down from the 3d to the 5d metals.

\subsection{Electronic structure : d-band center}
\label{Electronic structure : d-band center}

The position of the d-band center of the clean surface is another important characteristic which is closely related to the strength of the CO-surface interaction.  As it was already argued \cite{ss:Hammer:343,prl:Hammer:76,ac:Hammer:45}, the d-band centers play a significant role in the bonding for many adsorbate-substrate systems where the major interaction is due to  hybridization of the HOMO and LUMO of the adsorbate and the d-orbitals of the substrate.

\begin{figure}[htb]
\centerline{\psfig{figure=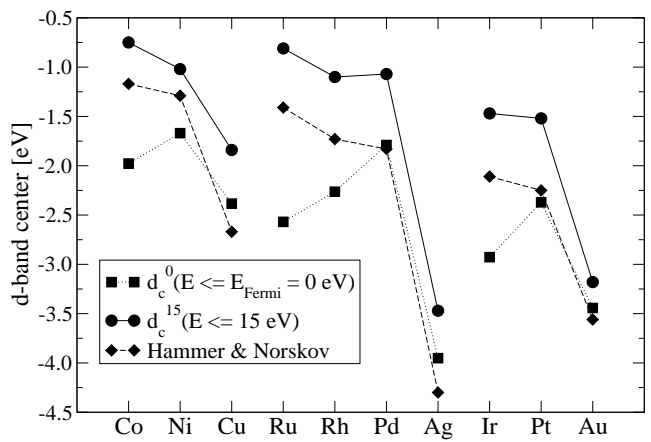,width=9.0cm,clip=true,angle=-90}}
\nopagebreak \caption{\label{figure dc clean} The d-band centers of the
clean TM surfaces integrated until E$_{\rm Fermi}$ level and 15~eV above E$_{\rm Fermi}$, where at
least 9.5 electrons are accommodated in the d-band. As a
comparison the calculated d-band centers from the article of
Hammer and  N{\o }rskov are taken \cite{ac:Hammer:45}.}
\end{figure}

However, simple as it might seen to be, the proper definition of the d-band center is not that straightforward:

(i) Within a plane-wave method, partial angular-momentum decomposed densities of states can be defined in different ways. We chose the projection of the plane-wave components onto spherical orbitals within some atomic radius.

(ii) DFT applies only to the ground-state, excited states are
usually predicted at too low energies. Therefore we report in Fig.
\ref{figure dc clean} the center of gravity of the occupied d-band
(integration up to the Fermi-level) d$_c^{0}$, and the center of
the entire d-band d$_c^{15}$ (integration from the lower band edge
up to 15~eV above E$_{\rm Fermi}$ where at least 9.5 electrons have been accommodated in
the d-band), together with the d-band centers taken from the older
LMTO calculations of Hammer and N{\o }rskov (d$_c^{HN}$)
\cite{ac:Hammer:45}. The d$_c^{0}$ and d$_c^{15}$ values provide
upper and lower bounds for our d-band centers. The d$_c^{HN}$ lies
in between for TM surfaces, but it is lower in energy for noble
metals. This stems from a small contributions to the density of
states far from the main d-band peak. As per definition, the
difference between d$_c^{0}$ and d$_c^{15}$ increases with
decreasing band-filling. d$_c^{15}$ shows only little variation
in groups VIII to IB. As expected, it drops significantly for the
noble metals compared to the metals in the same row.

\section{CO adsorption}
\label{CO adsorption}

We have studied the adsorption of CO  in the top, bridge and hollow (fcc, hcp) sites in the c($2\times4$) cell, corresponding to a coverage of a quarter of a monolayer ($\Theta$=0.25~ML) (see Fig.  \ref{figure structure} ). From these calculations we derive in the following
trends in the adsorption energies, the vibrational properties, the geometric and the electronic structure.

\subsection{Adsorption energy and site preference}
\label{Site preference}

Beside the adsorption geometry and the vibrational properties a precise determination of the adsorption energy for a given geometry is one of the keys to understanding the  mechanism of the catalytical activity and selectivity \cite{ss:Somorjai:89}.
The adsorption energy describes the strength of the chemical bond between the CO molecule and the metallic surface.

The adsorption energy for the stable adsorption site corresponds
to the absolute minimum on the potential energy surface (PES) of
the molecule moving on the surface. The adsorption energies
calculated for other high-symmetry sites correspond either to
local extreme or to a saddle point on the PES. The position and
the height of the saddle points define the diffusion path and the
activation energy for surface diffusion.

The experimental isosteric heat of the adsorption
H$_{\rm ads}$($\Theta$) is not directly comparable with the calculated
adsorption energy. If we want to compare our theoretically
calculated adsorption energy E$_{\rm ads}$ (i.e. the gain in energy
when a certain amount of CO is adsorbed) at $\Theta$=0.25 ML with
the isosteric heat of adsorption (a differential energy gained
when adsorbing one additional molecule on a surface at a certain
CO coverage) we need to take the integral up to the coverage used
in the calculation:

\begin{equation}
H_{\rm ads,int}(\Theta_{\rm calc})=\frac{1}{\Theta_{\rm calc}} \int_{0}^{\Theta_{\rm calc}}H_{\rm ads}(\Theta) d \Theta .
\end{equation}

The experimental heats of adsorption with and the  proposed adsorption sites are presented in Table \ref{table adsorption}. We show adsorption sites which are found at 0.25~ML or up to 0.5~ML coverage.

\begin{table*}[htb]
\begin{tabular}{rr|rrr|rrr|rrr}
\hline
\multicolumn{8}{c|}{Experiment} & \multicolumn{3}{c}{Theory} \\
\hline
surface & site   & H$_{\rm ads}$(0ML)&H$_{\rm ads}$($\frac{1}{4}$ML)& H$_{\rm ads,int}$ & method  & $\Theta$ [ML]  & ref. & E$^{PW91,450}_{\rm ads}$ & E$^{PW91,700}_{\rm ads}$ & E$^{RPBE,700}_{\rm ads}$\\
\hline
Co& top    & --1.33   & --1.33 &   --1.33   &SP & 0.33 & \cite{ss:Papp:129} &--1.65&--1.62&--1.32\\
  &        & --1.19   &        &            &TDS&      & \cite{ss:Lahtinen:418} & & &\\
\hline
Ni& hollow & --1.35   & --1.28  &   --1.32  &CAL &0.25&\cite{jcp:Stuckless:99}&--1.95&--1.90&--1.44\\
  &        & --1.31   & --1.28  &   --1.30  &TDS & 0.25&\cite{ss:Froitzheim:188}& & &\\
  &        & --1.31   & --1.06  &           &TDS & 0.25& \cite{jcp:Miller:87}   & & &\\
\hline
Cu& top    & --0.49   &        &            &TDS &     &\cite{cl:Vollmer:77}&--0.75&--0.72&--0.42\\
  &        & --0.425  &        &            &TDS &     & \cite{ss:Kirstein:176}  &  & &\\
  &        & --0.46   &        &            &TDS & 0.33& \cite{ss:Kessler:67} &  & & \\
\hline
Ru& top    &  --1.66  & --1.66 &   --1.66   &TDS & 0.20&\cite{jcp:Pfnuer:79} &--1.89 &--1.82&--1.69\\
\hline
Rh& top    & --1.5&--1.38&$\approx$--1.44&TDS&0.2-0.25&\cite{ss:Smedh:491:115}&--1.89&--1.86&--1.55\\
  &        & --1.37   & --0.77 &$\approx$--1.07&TDS & 0.25& \cite{ss:Thiel:84}& & &\\
  &        & --1.71   &        &              &TREELS&     & \cite{ss:Wei:381}       &  & & \\
  &        & --1.40   & --1.28 &  --1.36     &He  &     & \cite{jcp:Perlinz:95}   &  & & \\
\hline
Pd& hollow & --1.30   &        &              &TDS &    &\cite{jvst:Szanyi:11}&--2.14&--2.09&--1.68\\
  &        & --1.54   & --1.30 &$\approx$--1.42&TDS & 0.25& \cite{jcp:Guo:90}        & & &\\
\hline
Ag& top    &          &--0.28  &            &TDS &   & \cite{ss:McElhiney:54} &--0.16 &--0.14&0.18\\
\hline
Ir& top    & --1.81 &--1.55 &$\approx$--1.63&TDS&0.25&\cite{jvst:Sushchikh:15}&--1.98&--1.94&--1.64\\
  &        &          &--1.52  &            &   &0.33& \cite{jcp:Comrie:64}    &  & &\\
\hline
Pt& top    & --1.50   &        &            &TDS&0.25&\cite{ss:Steininger:123}&--1.70&--1.67&--1.34\\
  &        & --1.39   & --1.22 &$\approx$--1.33&LITD& 0.25& \cite{jvst:Seebauer:5}    & & &\\
  &        & --1.43   & --1.30 &$\approx$--1.37&TDS & 0.25& \cite{ss:Ertl:64}         & & &\\
\hline
Au& top    &         & --0.40  &              &TDS &     &\cite{Elliott}      &--0.32& --0.24 &0.12\\
\hline
\end{tabular}\caption{\label{table adsorption}
Experimental heats of adsorption for $\Theta$=0 and 0.25 ML (if
available) and the integral heat of adsorption between zero limit
and 0.25 ML. Integral heats of adsorption H$_{\rm ads,int}$ are
compared with our calculated adsorption energies E$_{\rm ads}$ at the
experimentally observed adsorption sites for the PW91 and RPBE
exchange-correlation functionals using two different sets of
pseudopotential (E$_{\rm cut}$~=~450~eV, E$_{\rm cut}$~=~700~eV, cf. text).
TDS - Thermal Desorption Spectroscopy, CAL - Calorimetry method,
He - Helium scattering method, LITD - Light Induced Thermal Desorption, SP - SP measurement, TREELS - Time Resolved Electron Energy Loss Spectroscopy. }
\end{table*}

The main trends in the experimental results given in Table \ref{table adsorption} include:
\begin{enumerate}
\item CO adsorption on TM surfaces is several times stronger (E$_{\rm ads}\sim$ 1.2 to 1.7~eV) than on noble metals (E$_{\rm ads}\sim$ 0.3 to 0.5~eV).
\item All measured adsorption sites (with the exception of on-top site on Ru(0001)) show a slight decrease of the heat of adsorption going from the zero coverage limit to a quarter-monolayer coverage.
\item The heat of adsorption decreases with increasing filling of the d-band.
\end{enumerate}

\begin{figure}[htb]
\centerline{\psfig{figure=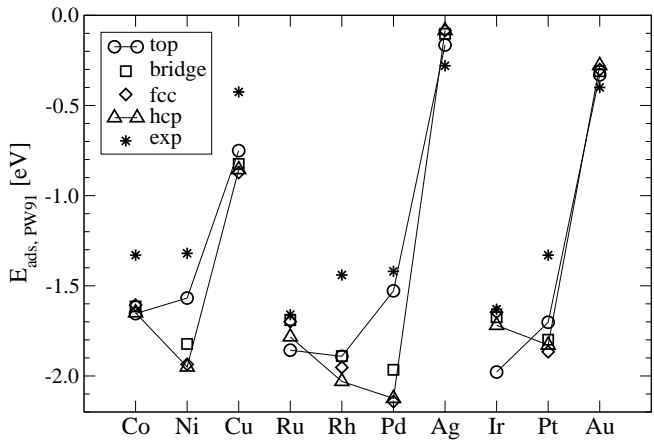,width=9.0cm,clip=true,angle=-90}}
\nopagebreak \caption{\label{figure adsorption PW91 450} Calculated CO adsorption
energies on TM surfaces for the PW91 exchange-correlation functional and energy cut-off 450~eV. Experimental CO heat of desorption are labelled by the stars (see Table \ref{table adsorption}).}
\end{figure}

Fig. \ref{figure adsorption PW91 450} summarizes the calculated
adsorption energies (using the PW91 GGA) for the high symmetry positions described in
Fig. \ref{figure structure}. For the transition metals we note a
pronounced tendency to overestimate the adsorption energies - this
is independent of an eventual relaxation of the surface. For the
noble metals (with the exception of Cu) on the other hand, the
calculations rather tend to underestimate the adsorption energies
- this could be related to the neglect of dispersion forces.

Experimentally, the adsorption energies on the TM surfaces are
found to decrease monotonously with increasing band filling. In
the calculation, the variation depends on the adsorption site:
E$_{\rm ads}$ decreases for on-top adsorption, but increases for sites
with higher coordination. This leads to wrong predictions for Cu,
Rh and Pt where the hollow sites were found to be preferred for CO
adsorption, whereas experiment shows that adsorption occurs on
the top sites. Moreover, the relaxation of
the metallic substrate is necessary for Co and Ru surfaces, i.e. if the surface is not allowed to relax we get a different site
prediction compared to the experiments. However, even after substrate relaxation, the energies for CO adsorption in hollow and top sites on Co, Ag, and Au surfaces remain almost degenerate.
Site preference is indeed a fundamental problem for CO adsorption
on metallic surfaces. We shall try to elucidate the origin of this
problem below.

Another interesting point is that among the hollow sites, the preference changes from hcp (Co, Ni, Ru, Rh, Ir) to fcc (Cu, Pd, Ag, Pt, Au) with increasing d-band filling.

\subsection{Geometric structure}
\label{Geometric structure}

In the past many studies have demonstrated that on the basis of
DFT a  reliable determination of the geometrical structure is
possible. The local adsorption geometry of the CO molecule on an
unrelaxed TM surface can uniquely be characterized by the
S(urface)-C distance (height above the surface) and the C-O bond
length. If additionally the metal atoms are allowed to relax, this
modification of the surface can be described by the change of the
inter-layer distance and the buckling of the surface.

On all studied surfaces the CO molecule adsorbs in an upright
position with the carbon atom pointing towards the surface. The
tilt with respect to the surface normal is always less than
2$^{\circ}$.

\begin{figure}[htb]
\centerline{\psfig{figure=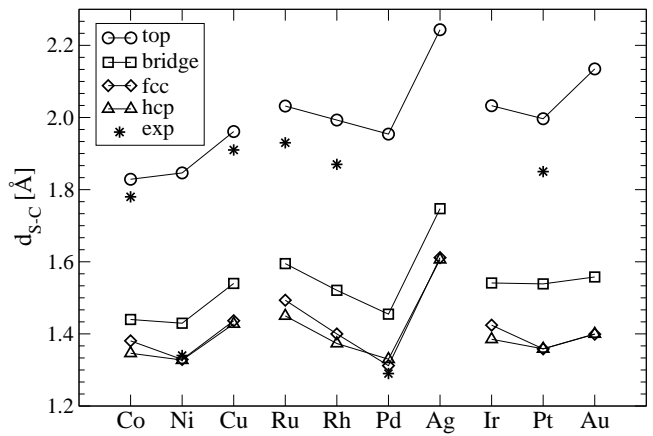,width=9.0cm,clip=true,angle=-90}}
\nopagebreak \caption{\label{S-C length} Calculated heights of the CO
adsorbate (surface-C distance) in different adsorption sites on various TM surfaces together with values taken from experimental literature (Table \ref{table structure}).}
\end{figure}

The minimum distance between neighboring adsorbates varies in our setup
(c(2$\times$4)) between 4.3~\AA~(Co) to 5.1~\AA~(Au). Therefore direct
as well as indirect interactions can be considered
to be small and our calculations probe the low-coverage case.

In response to the adsorption of the CO molecule, the metal atoms
in the immediate surrounding move outwards. This effect is quite localized.
The buckling is less
pronounced for adsorption in higher coordinated sites ($\sim$~0.1~\AA) and
increases for one-fold coordinated CO in an on-top position to $\sim$0.2~\AA.

The height of the CO molecule above the surface (d$_{\rm S-C}$) is
determined by the extension of the metal d-orbitals and (for the
higher coordinated sites) by the nearest neighbor distance of the
substrate atoms. The greater the extension of the interacting
metal orbitals the larger is the distance between the C-atom and the closest metal atom (d$_{\rm M-C}$).
Similarly, for a given metal-C distance, the molecule will adsorb at
lower height for greater lattice constants. The essential importance of the metal-carbon bond distances
for understanding the mechanism of the metal-CO bonding was
pointed out earlier \cite{ss:Ishi:161}.

\begin{figure}[htb]
\centerline{\psfig{figure=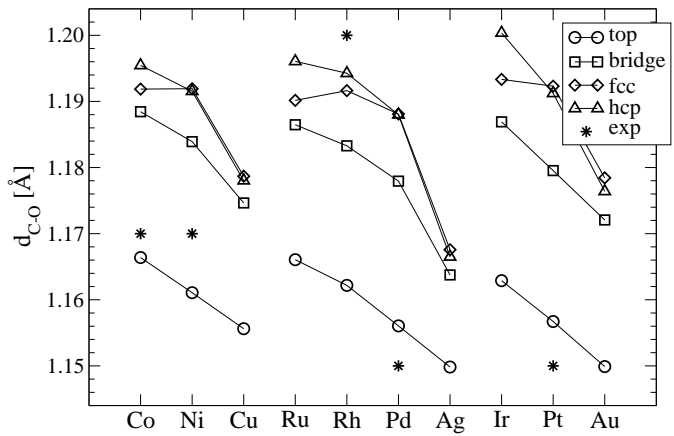,width=9.0cm,clip=true,angle=-90}}
\nopagebreak \caption{\label{C-O bond length} Calculated C-O
bond lengths in top, bridge, fcc and hcp hollow sites on TM surfaces together with experimental values as listed in Table \ref{table structure}.}
\end{figure}

As it was already indicated in the previous studies, a different occupation of the interacting d-band orbitals leads to different values for the d$_{\rm S-C}$ and d$_{\rm M-C}$ lengths.
In Figs. \ref{S-C length} and \ref{C-O bond length} the main characteristics of the CO adsorption
geometry on TM surfaces are compiled.

When comparing the adsorption geometry, the noble metals behave
differently from all other metals. This reflects the completely
filled d-band and hence the different bonding mechanism.
Adsorption on the noble metals is significantly weaker (see
adsorption energies Fig. \ref{figure adsorption PW91 450}), and
consequently the d$_{\rm S-C}$ and d$_{\rm M-C}$ lengths are enhanced (see
Fig. \ref{S-C length}). On the other hand, CO adsorption on the
surfaces of TM is stronger and d$_{\rm S-C}$ (resp. d$_{\rm M-C}$) tends
to decrease as the filling of the d-band increases. This trend is
more pronounced for 4d than for 5d or 3d metals (for Co and Ni
this does not hold for on-top adsorption). For on-top
adsorption, the height of the adsorbate increases going from 3d
via 4d to 5d elements. The coordination has a strong effect on the
CO height: CO in the lower coordinated sites is closer to the
surface.

The C-O bond length (d$_{\rm C-O}$) of the free molecule was calculated to be 1.142~\AA. On adsorption, the d$_{\rm C-O}$ length increases with the coordination of the adsorption site, similarly to d$_{\rm S-C}$ and d$_{\rm M-C}$: d$_{\rm C-O}$ rises from $1.15-1.17$~\AA\ for one-fold to $1.17-1.20$~\AA\ for three-fold coordinates sites. This trend was already stressed in the experimental work of Westerlund et al. \cite{ss:Westerlund:199}. Additionally, the C-O bond is more elongated as the filling of the d-band decreases.

\begin{table*}[htb]
\begin{tabular}{rrrrrr|rr}
\hline
\multicolumn{6}{c|}{Experiment} & \multicolumn{2}{c}{Theory} \\
\hline
surface & site & d$_{\rm S-C}$~(\AA) & d$_{\rm C-O}$~(\AA) & $\Theta$~(ML)  & ref. & d$^{cal}_{\rm S-C}$~(\AA) & d$^{cal}_{\rm C-O}$~(\AA)\\
\hline
Co  & top       & 1.78$\pm$0.06     & 1.17$\pm$0.06     & 0.33         & \cite{ss:Lahtinen:448}   & 1.83 & 1.166\\
\hline
Ni  & hollow & 1.34$\pm$0.07     & 1.15$\pm$0.07     & 0.5            & \cite{ss:Eichler:526}      & 1.329 &  1.192\\
       & hollow & 1.29$\pm$0.08     & 1.18$\pm$0.07     & 0.5            & \cite{ss:Eichler:526}      & 1.327  & 1.192\\
\hline
Cu  & top       & 1.91$\pm$0.01     &                    &0.33,0.44&  \cite{prb:Moler:54}       & 1.96  & 1.156\\
\hline
Ru  & top       & 1.93$\pm$0.04     & 1.10$\pm$0.05     & 0.33           & \cite{prl:Over:70}          & 2.03  & 1.166\\
\hline
Rh  & top       & 1.87$\pm$0.04     & 1.20$\pm$0.05     & 0.33           & \cite{ss:Gierer:391}      & 1.99  & 1.162\\
\hline
Pd  & hollow & 1.27$\pm$0.04     & 1.14$^{+0.14}_{-0.11}$& 0.33     & \cite{ss:Giessel:406}
& 1.31 &  1.188\\
    & hollow & 1.29$\pm$0.05     & 1.15$\pm$0.04         & 0.33     & \cite{ss:Ohtani:187}       & 1.33 &  1.188\\
\hline
Pt  & top    & 1.85$\pm$0.10     & 1.15$\pm$0.1          & 0.3      & \cite{ss:Blackman:61}
& 2.00   & 1.157\\
\hline
\end{tabular}
\caption{\label{table structure}
Experimental geometrical structures from low energy electron diffraction (LEED) studies of the CO adsorption on TM surfaces together with the values obtained in this study (superscript ``cal'').
The normal distance between carbon and the TM surface (d$_{\rm S-C}$) and C-O bond length (d$_{\rm C-O}$) at the coverage $\Theta$ together with corresponding references.}
\end{table*}

There are many data for the geometrical structure of the CO adsorption on TM from past experiments \cite{ss:Lahtinen:448}-\cite{ss:Blackman:61}. Only for iridium, gold and silver surfaces we are not aware of any experimental literature. A valuable collection of the detailed experimental data for the adsorption of small molecules on the metallic surfaces can be found in the study of Over \cite{pss:Over:58}. A part of the vast data is presented in Table \ref{table structure}.

The main difficulty in comparison with experiment (LEED, XPD,
EXAFS, ...) is that the estimated error for the  C-O bond length
($\sim\pm$0.05\AA) is mostly too large, covering the whole interval of
all calculated d$_{\rm C-O}$ values. Although the c(2$\times$4)-CO
structure used in the calculation is not realized on all metallic
surfaces, values for p($\sqrt{3}\times\sqrt{3}$) and p($2\times2$)
structures give values in line with our results. This implies that
the CO adsorbate-adsorbate interactions at a coverage lower than
$\Theta$~=~$\frac{1}{3}$~ML play only a minor role.

\subsection{Vibrational frequencies}
\label{Vibrational frequencies}

Vibrational frequencies can be measured very accurately using infrared spectroscopy (RAIRS), electron energy loss spectroscopy (EELS) and sum frequency generation (SFG) methods. The dependence of the stretching frequency $\nu_{\rm C-O}$ on the coordination by the surface atoms has often been used as an indication of the adsorption site. The correlation between the vibrational and electronic properties was analyzed in the paper by Ishi et al. \cite{ss:Ishi:161} in which a correlation between $\nu_{\rm C-O}$ and the energy level difference between $5\tilde{\sigma}$ ($5\sigma$ character after adsorption) and $1\tilde{\pi}$ orbitals of the adsorbed CO ($\Delta(5\tilde{\sigma}-1\tilde{\pi})$) was suggested.

\begin{figure}[htb]
\centerline{\psfig{figure=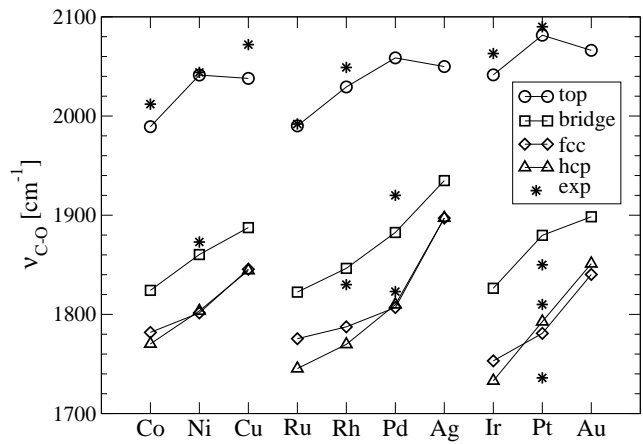,width=9.0cm,clip=true,angle=-90}}
\nopagebreak \caption{\label{figure C-O stretch} Calculated C-O stretching frequencies ($\nu_{\rm C-O}$) on closed-packed TM surfaces together with experimental values as listed in Table \ref{table frequency}. The calculated value of $\nu_{\rm C-O}$ for the free CO molecule is 2136~cm$^{-1}$ compared to the experimental value of 2145~cm$^{-1}$.}
\end{figure}

\begin{figure}[htb]
\centerline{\psfig{figure=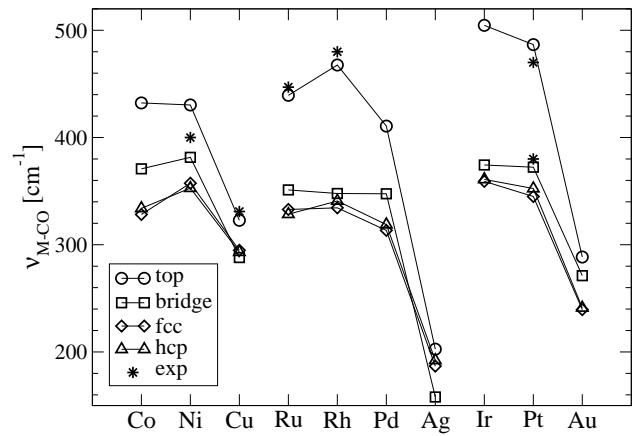,width=9.0cm,clip=true,angle=-90}}
\nopagebreak \caption{\label{figure M-C stretch} Calculated M-CO vibrational
frequencies ($\nu_{\rm M-CO}$) on closed packed TM surfaces together with experimental values as compiled in Table \ref{table frequency}. }
\end{figure}

 In the following we will focus on two eigenmodes: the M-CO ($\nu_{\rm M-CO}$) and the C-O
($\nu_{\rm C-O}$) stretching frequencies. Our calculated $\nu_{\rm C-O}$ values
are compiled in Fig. \ref{figure C-O stretch}. For free CO
molecules a stretching frequency of $\nu_{\rm C-O}$ 2136~cm$^{-1}$ was calculated,
this value is lowered due to the formation of the
bond between the molecule and the metallic surface. The C-O
vibrations exhibit the expected strong coordination dependence: CO
molecules adsorbed in the lower coordinated sites vibrate faster,
typical values for the C-O stretching frequency are
1990--2100~cm$^{-1}$, 1830--1880~cm$^{-1}$, 1750--1810~cm$^{-1}$\
for carbon monoxide in top, bridge and hollow sites. The
stretching frequency for the CO molecule in bridge and hollow
sites on noble metals is higher than on the TMs whereas
for on-top adsorption the frequencies are almost the same for the
noble metals and those of the Pt group.

The dependence of $\nu_{\rm C-O}$ on the d-band filling of the substrate is also obvious: increased d-band filling raises $\nu_{\rm C-O}$; the correlation is essentially linear. An exception to the rule is CO in on-top sites on noble metals. With decreasing d-band filling the stretching frequencies for the CO adsorbed in the fcc and hcp hollows begin to differ, with the lower $\nu_{\rm C-O}$ for the hcp sites.

The adsorbate-substrate stretching frequencies $\nu_{\rm M-CO}$  are compiled in Fig. \ref{figure M-C stretch}. $\nu_{\rm M-CO}$ exhibits a similar coordination dependence as $\nu_{\rm C-O}$: CO molecules adsorbed in the lower coordinated sites vibrate faster, typical intervals of the $\nu_{\rm M-CO}$ stretching frequencies on the TM surfaces are 400--475~cm$^{-1}$, 340--380~cm$^{-1}$, 300--350~cm$^{-1}$ for top, bridge and hollow sites. There is no pronounced correlation with the d-band filling.  For noble metals $\nu_{\rm M-CO}$ lies generally below 300~cm$^{-1}$ and there is only a very weak site dependence. Once more, this reflects the weak CO-noble metal bonding. The metal-substrate stretching frequency is particularly low for the Ag surface, consistent with a very low adsorption energy. We have collected available experimental data for $\nu_{\rm C-O}$ and $\nu_{\rm M-CO}$ frequencies in Table \ref{table frequency} and included them in Fig. \ref{figure C-O stretch} and \ref{figure M-C stretch}.  Agreement between the experiment and theory is very good.

\begin{table*}[htb]
\begin{tabular}{rrrrrrr|rr}
\hline
\multicolumn{7}{c|}{Experiment} & \multicolumn{2}{c}{Theory} \\
\hline
surface  & site  & $\nu_{\rm M-CO}$~(cm$^{-1}$) & $\nu_{\rm C-O}$ (cm$^{-1}$) &  $\Theta$ (ML) & Method & Ref. & $\nu^{cal}_{\rm M-CO}$ & $\nu^{cal}_{\rm C-O}$ \\
\hline
Co& top    &        & 2012       &   0.33         &RAIRS&   \cite{jpc:Beitel:100}     &432&1989\\
\hline
Ni& top    &        & 2044       &   0.25         &RAIRS,HREELS&   \cite{ss:Eichler:526} &430&2041\\
  & hollow &  400   & 1873       &   0.25         &RAIRS,HREELS&   \cite{ss:Eichler:526} &353&1804\\
\hline
Cu& top    &   346  & 2072       &                &RAIRS&   \cite{jesrp:Hirschmugl:54} &323 &2038\\
  &        &   331  & 2077       &   0.33         &EELS,RAIRS&   \cite{ss:Raval:203}   & - & - \\
  &        &        & 2075       &   low(0.5L)    &RAIRS&   \cite{cpl:Eve:313}         & - & - \\
\hline
Ru& top    &   445  & 1980-2080  &   0.07         &EELS&   \cite{jcp:Thomas:70}    & 439 & 1990 \\
  &        &   447  & 1990       &   0.33         &RAIRS&   \cite{prl:Jacob:78}    & - & - \\
  &        &        & 1992       &   0.25         &HREELS&   \cite{ss:He:345}         &-&-\\
\hline
Rh& top    &        & 2000       &   0.33         &HREELS&   \cite{ss:Gierer:391}   & 468 & 2029 \\
  &        &   480  & 1990       &   low          &HREELS&   \cite{ss:Dubois:91}    & - & - \\
  &        &        & 2049       &   0.2          &HREELS&   \cite{ss:Smedh:491:99} & - & - \\
  & hollow &        & 1830       &   0.2          &HREELS&   \cite{ss:Smedh:491:99} & 328 & 1782 \\
\hline
Pd& hollow &        & 1848       &    0.33        &RAIRS& \cite{ss:Giessel:406}   & 319 & 1810 \\
  &        &        & 1823       &   $<$0.25      &RAIRS& \cite{ss:Bradshaw:72}& -&- \\
  &        &        & 1823-1850  &   $<$0.33      &HREELS& \cite{ss:Surnev:470}   & -&- \\
  & bridge &        & 1920       &    0.5         &RAIRS&  \cite{ss:Giessel:406}  & 348 & 1883 \\
\hline
Ag& top    &        & 2137       &    low(2L)     &HREELS&    \cite{ss:Hansen:253}  & 203 & 2050 \\
\hline
Ir& top    &        & 2063       &    low(1L)        &RAIRS&   \cite{jvst:Sushchikh:15}&505& 2041\\
  &        &        & 2028-2090  &    0-0.71ML    &RAIRS&    \cite{jvst:Schick:14}    & -&- \\
  &        &        & 2065       &     0.25       &RAIRS&    \cite{ss:Lauterbach:350} & -&- \\
\hline
Pt& top    &    470 & 2100       &     0.24       &EELS&   \cite{ss:Steininger:123}   & 487& 2081\\
  &        &        & 2090       &                &SFG&   \cite{ss:Kung:463}          & -& -\\
  &        &    464 &            &     0.5        &RAIRS&   \cite{ss:Surman:511}      & -& -\\
  &        &    467 & 2104       &     0.5        &RAIRS&   \cite{ss:Schweizer:213}   & -& -\\
  &        &        & 2093       &     0.07       &RAIRS&   \cite{cp:Nekrylova:205}   & -& -\\
  &        &        & 2095       &     0.1        &RAIRS&   \cite{ss:Yoshinobu:363}   & -& -\\
  &        &        & 2100       &     0.25       &RAIRS&   \cite{ss:Heyden:125}      & -& -\\
  & bridge &  380   & 1850       &     0.24       &EELS&   \cite{ss:Steininger:123}   & 372&1880\\
  &        &  376   &            &     0.5        &RAIRS&    \cite{ss:Surman:511}     & -& -\\
  &        &        & 1855       &     0.5        &RAIRS&    \cite{ss:Schweizer:213}  & -& -\\
  &        &        & 1858       &     0.07       &RAIRS&   \cite{cp:Nekrylova:205}   & -& -\\
  & hollow &        & 1810       &     0.5        &RAIRS&   \cite{ss:Heyden:125}      & 352& 1793\\
  &        &        & 1736       &     0.51       &RAIRS&   \cite{ssl:Nekrylova:295}  & -& -\\
\hline
\end{tabular}
\caption{\label{table frequency} Experimental and calculated
metal-CO ($\nu_{\rm M-CO}$) and C-O vibrational frequencies
($\nu_{\rm C-O}$) of CO adsorbed on  3d and 4d TM surfaces. The
coverage $\Theta$ and references to the experimental studies are
added. RAIRS - Reflection-Absorption Infra-Red Spectroscopy,
(HR)EELS - (High Resolution) Electron Energy Loss Spectra, SFG -
Sum Frequency Generation spectroscopy. }
\end{table*}

As a final remark we emphasize that here only high symmetry sites have been considered. A mixed occupation of different sites or off-symmetry adsorption at higher coverage can significantly increase the CO stretch frequency. As an example we refer to a recent publication on adsorption of CO on the Ni(111) surface presented by Eichler \cite{ss:Eichler:526}.

\subsection{Electronic structure}
\label{Electronic structure}

The electronic structure provides deep insight into the interaction between adsorbate and surface. There are many papers which deal with electronic structures and their influence on the trends in binding energy or geometrical structure of the CO molecule on metallic surface \cite{ss:Ishi:161,ac:Hammer:45}.
The importance of the interplay between the geometric and the electronic structure in the understanding of CO adsorption was stressed by F\"ohlisch et al. \cite{prl:Foehlisch:85}. In our analysis we essentially follow the ideas expressed in Refs. \cite{prb:Kresse:68,ss:Bagus:278,rmp:Hoffmann:60}.

How is CO adsorbed on the TM surface ? It is generally assumed that a major part of the CO - metal interaction can be explained in terms of frontier orbitals (the highest occupied molecular orbital (HOMO) and the lowest occupied molecular orbital (LUMO) orbitals). The Blyholder model is based on the donation from the occupied  CO-$5\sigma$ states into empty surface orbitals and the back-donation from occupied surface orbitals to the CO-$2\pi^{\star}$ orbitals \cite{jpc:Blyholder:68}.

At this point we can divide the literature into three groups. To
the first belong those that agree with the Blyholder model
completely \cite{jpc:Blyholder:68,ss:Illas:376,ss:Aizawa:399}. The
second group consists of those that agree with Blyholder (in
principle), but point out that the Blyholder model ignores the
contribution to the bonding from the 4$\sigma$ and $1\pi$
orbitals. Some consider the model not only oversimplified, but
propose another $2\pi^{\star}$ resonance model like Gumhalter et
al.  \cite{prb:Gumhalter:37}. And finally to the third group
belong those that disagree with the Blyholder model and claim that
there is no back-donation to the $2\pi^{\star}$ orbital
\cite{prb:Ohnishi:49}.

\begin{figure*}[phtb]
\centerline{\psfig{figure=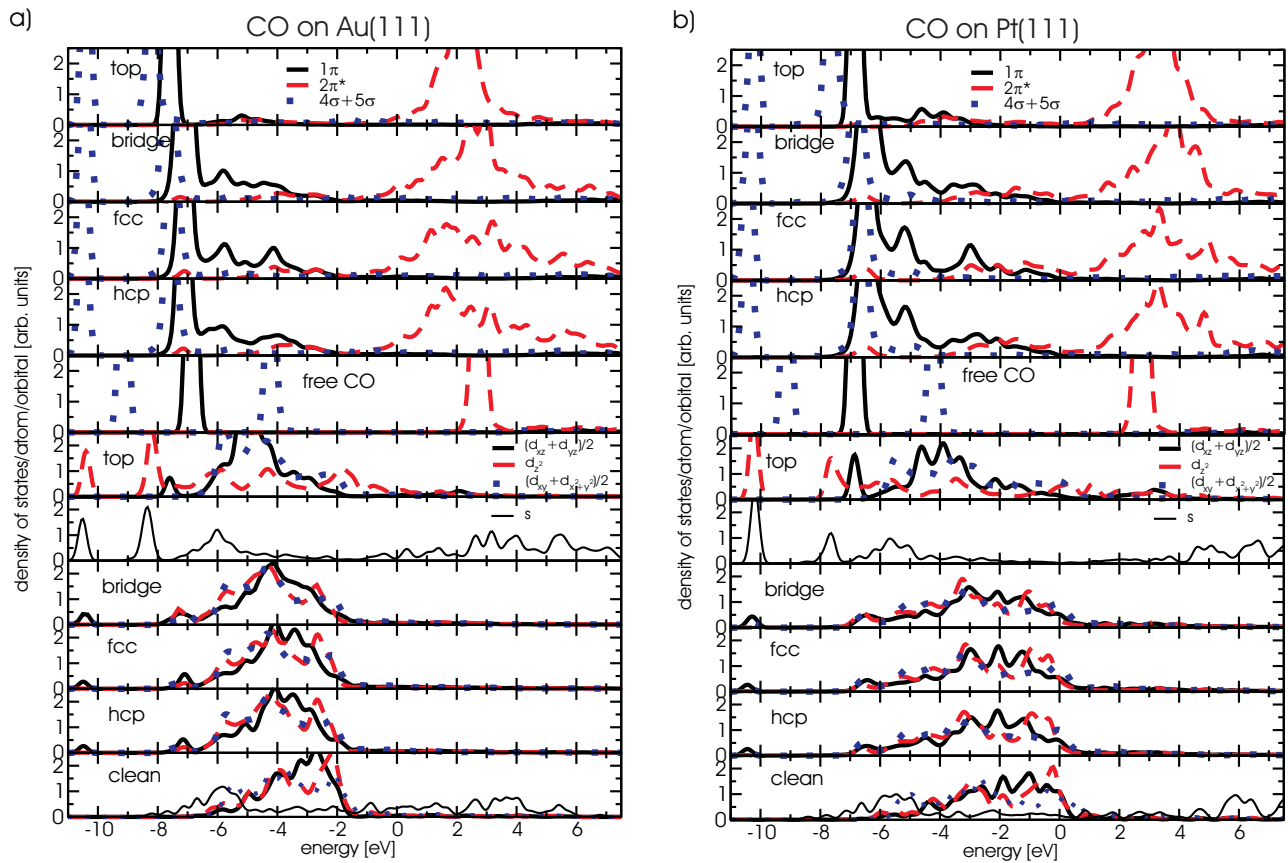,width=17.0cm,clip=true}}
\caption{\label{figure CO Au} The projected electronic densities of states (PDOS) for the CO molecule adsorbed in top, bridge, fcc and hcp hollow sites of the Au (a) and Pt (b) surface. While the upper five panels describe the PDOS for various molecular orbitals of the CO molecule, the lower six panels show the PDOS for the substrate atom(s) interacting most with the molecule. The panels labelled 'free CO' and 'clean' describe the noninteracting case for comparison (molecule 4~\AA ~above the surface). The DOS is smoothed by a Gaussian function
 with a width of 0.2~eV and the Fermi level is located at 0~eV. }
\end{figure*}

 Although in the experimental study of Nilsson et al. \cite{prl:Nilsson:78}, the authors emphasize that an atom-specific look at the CO adsorption could provide insight into the surface chemical bond, we have projected the DOS onto the molecular CO orbitals.
The density of states projected onto $1\pi$, $2\pi^{\star}$, $5\sigma$ and $4\sigma$ molecular orbitals (PDOS) of a CO molecule in top, bridge, hollow sites and a CO molecule far above ($\sim$4 \AA) the surface is depicted in the Fig \ref{figure CO Au}.
The weakly bonded CO molecules on the Au(111) surface already render the general interaction trends. The corresponding analysis of the orbitals of a CO molecule on Pt(111) can also be found in the paper by Kresse et al. \cite{prb:Kresse:68}.

 There is no principal difference between the Au and Pt DOS. As the CO molecule approaches the surface the localized CO orbitals, $3\sigma$ (not shown) and $4\sigma$, are shifted to lower energies, depending on the coordination between $\sim$1
and 3 eV. Since they are fully occupied they play only a minor role in the interaction with the metal, characterized
as Pauli repulsion. On the other hand, the 5$\sigma$, 1$\pi$ and
$2\pi^{\star}$ peaks broaden and dominate the interaction. Again,
the broadening and the shift of the peaks increase with
coordination: the higher the coordination - the greater the shift
and the broadening. This behavior is more visible for the
5$\tilde{\sigma}$ than for the $2\tilde{\pi}^{\star}$ peak. The
position of the $5\tilde{\sigma}$ band peak is almost always lower
than the $1\tilde{\pi}$ peak which is shifted by $\sim$1 to 3~eV.
The $5\tilde{\sigma}$ band is higher in energy only for on-top
adsorption of CO on the Ag(111) surface, which is partially
related to the weak interaction. A typical shift for the 5$\sigma$
peak is 3 to 4 eV, depending on the metal CO bond strength.
Finally, also the $2\pi^{\star}$ band broadens with increasing
coordination and is therefore partially shifted below the Fermi
level (E$_{\rm Fermi}$ = 0 eV) as shown in Fig \ref{figure CO Au}.

The detailed interaction picture on different sites depends on the
symmetry of  the surface and adsorbate orbitals. For the top site
the major orbital interaction (due to symmetry) is $5\sigma$ (CO)
- d$_{\rm z^2}$ (metal), whereas for the higher coordinated sites the
$1\pi$ and $2\pi^{\star}$ CO molecular orbitals are more important, interacting with the
d$_{\rm xz}$ (resp. d$_{\rm yz}$) and in-plane (d$_{\rm x^2-y^2}$, d$_{\rm xy}$)
orbitals of the metal atoms. The 5$\sigma$ orbital of the CO orbital
adsorbed on-top hybridizes with the d$_{\rm z^2}$ states of the
metallic substrate and shifts to lower energies. The d$_{\rm z^2}$
states broaden and split into a  $5\sigma-d_{\rm z^2}$  bonding
contribution far below the Fermi-level ($\sim$ 7.5 eV) and
$5\sigma-d_{\rm z^2}$ anti-bonding contributions located above the
bonding peak and partly even above the Fermi-level. This interaction
would be repulsive (Pauli like) if the $5\sigma-d_{\rm z^2}$ were not
pushed partly above the Fermi-level. A $5\sigma$ depletion
(donation from CO to metal) is in accordance with the Blyholder
argument of donation of electrons from the adsorbate to the
surface \cite{jpc:Blyholder:68}. The interaction between the 5$\sigma$
orbitals and the metal s-band is attractive, but depends on the amount
of electrons accumulated in the newly created molecular orbitals
of the CO-metal system. If we consider a depletion of the
5$\sigma$ orbital, then the interaction with the metallic s-band becomes more attractive.

The interaction between the $1\pi$(resp. $2\pi^{\star}$) orbitals and the substrate is more complex \cite{prl:Foehlisch:85}; we can distinguish four contributions:
(i) The main part of the $1\pi$ CO orbital is located at energies around $\sim-$6 eV. (ii) During the adsorption the peak broadens at the higher energy end up to the Fermi level. This broad state is usually called d$_{\rm \tilde{\pi}}$ orbital. All the interactions with the $1\pi$ orbital are located below the Fermi level and have repulsive character \cite{cpl:Hu:246}. (iii) Furthermore, in the same energy region is a contribution from a partially occupied $2\pi^{\star}$ orbital which increases with coordination and decreases with the d-band filling (see Fig. \ref{figure CO Au}). This contribution develops into a broad peak for Pt.
(iv) Finally, at higher energies ($\sim$3~eV) the anti-bonding 2$\pi^{\star}$-d$_{\rm yz}$(d$_{\rm xz}$) hybridized orbital is visible.

The later two contributions can be characterized by the fractional occupation of the $2\tilde{\pi}^{\star}$ orbital, compiled in Table
\ref{table 2pi occupation}.
At this point one should remark that our representation of the DOS
projected onto molecular orbitals is subject to small inaccuracies
not in position, but in amplitude arriving from the projection of
the plane-wave components. There exists a tiny hybridization between the molecular $1\tilde{\pi}$ and
$2\tilde{\pi}^{\star}$ orbitals, therefore the
calculated DOS is slightly overestimated at the position of the
$2\tilde{\pi}^{\star}$ for the $1\tilde{\pi}$ and at the position
of the $1\tilde{\pi}$ for $2\tilde{\pi}^{\star}$. This must be
considered when calculating the occupation of the
$2\tilde{\pi}^{\star}$ orbital.

\begin{table}[htb]
\begin{tabular}{r|rrr|rrrr|rrr}
\hline
\multicolumn{11}{c}{Occupation of 2$\tilde{\pi}^{\star}$} \\
\hline
site & Co &  Ni & Cu & Ru & Rh & Pd & Ag & Ir & Pt & Au \\
\hline
top   & 16.0  & 11.5  &  8.1  & 11.4  & 10.8  &  8.8  &  5.3  & 11.4  & 10.0  &  6.0 \\
bridge& 19.5  & 14.5  & 11.1  & 12.5  & 11.4  & 10.5  &  7.0  & 11.1  & 10.1  &  7.4 \\
fcc   & 29.9  & 22.1  & 16.3  & 18.5  & 18.5  & 18.1  & 11.7  & 20.5  & 19.8  & 15.0 \\
hcp   & 29.7  & 21.6  & 16.1  & 19.9  & 19.5  & 18.2  & 11.5  & 21.4  & 19.7  & 14.4 \\
\hline
\end{tabular}\caption{\label{table 2pi occupation}
Fractional occupation of the $2\tilde{\pi}^{\star}$ orbital for the CO molecule in top, bridge, fcc and hcp sites on the close-packed transition metal surfaces in \%.
}
\end{table}

The occupation of the $2\tilde{\pi}^{\star}$ orbital increases with coordination, but decreases with d-band filling and also from 3d to the 4d and 5d elements. We can further correlate the site preference between the two competitive (fcc and hcp) hollow sites to the occupation of $2\tilde{\pi}^{\star}$ orbital. As the occupation of the orbital for the fcc site compared to hcp site increases, the site preference changes from the hcp to the fcc hollow sites. This effect is probably due to a better interaction with the metal atom in the second layer below the hcp site for metals with smaller lattice constants.

\subsection{Charge density redistribution}

\begin{figure}[phtb]
\centerline{\psfig{figure=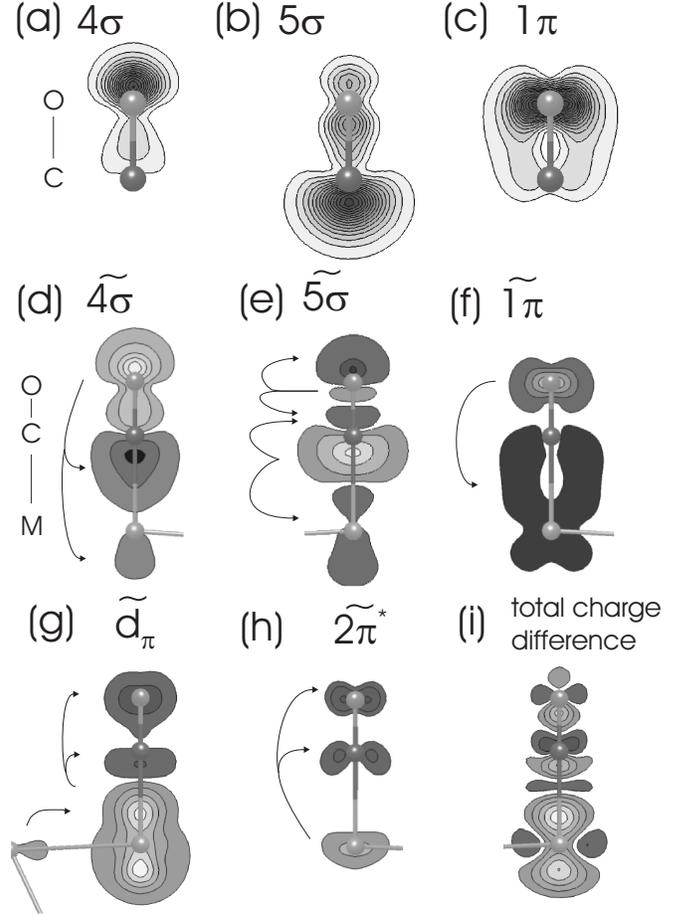,width=9.0cm,clip=true}}
\nopagebreak
\caption{\label{figure charge CO Au} Charge density of
(a) 4$\sigma$, (b) 5$\sigma$, (c) 1$\pi$ CO orbitals for the free
molecule and the difference
($\Delta\rho=\rho_{\rm CO+Au(111)}-\rho_{\rm Au(111)}-\rho_{\rm CO}$) after
the adsorption in the top site of the Au(111) surface (d, e, f).   Figures
(g) and (h) show the change in the metal charge density due to CO adsorption in the energy interval ($-$5.9, $-$1.0) and just below Fermi level
($-$0.2, E$_F$). The total charge density difference due to
adsorption is shown in Fig. (i). Dark regions: charge accumulation, light regions: charge depletion.}
\end{figure}

In this section, we illustrate the ideas derived in the previous
section from the analysis of the DOS by analyzing the charge
density and the difference in the charge density for the molecular
orbitals of the CO molecule adsorbed on-top on Au(111) surface as
shown in Fig. \ref{figure charge CO Au}. At first we show the
4$\sigma$, 5$\sigma$ and 1$\pi$ orbital-decomposed charge density
of the free CO molecule (panels a--c). We can see the accumulation
of the charge density for the 4$\sigma$ and 1$\pi$ orbitals around
the oxygen atom and for the 3$\sigma$ (not shown) and $5\sigma$
orbitals on the carbon atom. Fig. \ref{figure charge CO Au}d,
\ref{figure charge CO Au}e and \ref{figure charge CO Au}f
illustrate what happens to these orbitals after the CO molecule is
chemisorbed on-top on the Au(111) surface. Within the $4\sigma$
orbital charge density moves from the C-O bond and from above the
oxygen atom to the region below the carbon (Fig. \ref{figure charge CO Au}d).
Fig. \ref{figure charge CO Au}e shows
redistribution of charge in the $5\sigma$ molecular orbital:
charge is transferred from the side located closer to the metal to
the opposite side of the carbon and oxygen atom. This charge
transfer is in line with the experimental observation of a lone
pair localized on the oxygen atom as reported in Ref.
\cite{prl:Nilsson:78}. There is only little transfer of
charge density from the oxygen orbitals with $p_x/p_y$ character
into the metal-carbon bond ($1\tilde{\pi}$). This re-distribution
could be related to the slight inaccuracy of the orthogonalization
of $1\tilde{\pi}$ and $2\tilde{\pi}^{\star}$ orbitals.

The newly formed hybrid states, either by broadening of the $1\pi$
orbital (d$_{\rm \tilde{\pi}}$) or by occupation of 2$\pi^{\star}$ orbitals is shown in
panels g and h. The d$_{\rm \tilde{\pi}}$ orbital shows a charge transfer
from the d-orbitals towards the $\pi$-like molecular orbitals. We can
even spot a small depletion of charge at the metal atom which is
only a 2nd nearest neighbor of the CO, charge which is attracted towards the metal atom interacting with
the CO molecule. Furthermore, $2\tilde{\pi}^{\star}$ states become
occupied as can be seen in Fig. \ref{figure charge CO Au}h. In the
panel i, we present the total charge density difference of all
states, i.e. the sum of all mentioned charge density
redistributions.

Summarizing, we can say that there is a depletion
of charge in the substrate out-of-plane d-orbitals and
accumulation of charge density in the in-plane d-orbitals of the
substrate. The charge density on CO is re-distributed and
additionally some charge is transferred into $2\pi^{\star}$-like orbitals,
which weakens the C-O bond. The establishment of the M-CO bond is
reflected by the accumulation of charge between the bonding
species.

\section{Improving site preference}
\label{Improving Site Preference}

\begin{figure}[phtb]
\centerline{\psfig{figure=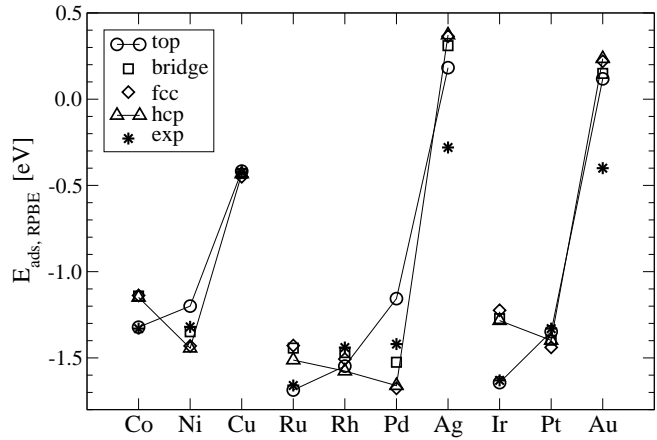,width=9.0cm,clip=true,angle=-90}}
\nopagebreak \caption{\label{figure RPBE 700} CO adsorption
energies on TM surfaces for the RPBE exchange-correlation functional and energy cut-off 700~eV. Experimental CO heat of desorption are labelled by the stars (see Table \ref{table adsorption}), calculated adsorption energies for top sites are presented with triangles, bridge - squares and fcc hollow - diamonds and hcp sites with circles.}
\end{figure}

As previously mentioned the calculations for Cu, Rh, Pt surfaces
predict a wrong site preference for CO adsorption and the adsorption energies for different adsorption sites on Co, Ag and Au are almost degenerate, so that small changes in the setup (k-points and neglect of surface relaxation) can alter the site preference. Can harder pseudopotentials (lower r$_{\rm cut}$ in the pseudopotential generation) or
different GGA approaches influence the site preference ? In order to shed light onto this question
we repeated our calculations performed with the PW91 exchange-correlation
functional and cut-off energy E$_{\rm cut}$~=~450~eV  with a harder pseudopotential
which requires a higher E$_{\rm cut}$~=~700~eV and used two different
GGA functionals: PW91, RPBE.  The use of RPBE is known to result in lower adsorption energies for CO, improving the agreement with experiment in most cases \cite{prb:Hammer:59} at almost unchanged geometrical structure. The results for RPBE with E$_{\rm cut}$~=~700~eV (calculated at the optimized PW91 geometry) are presented in Fig. \ref{figure RPBE 700}.

Switching to the RPBE GGA changes the energetics drastically by $\sim$0.3 eV and
the values agree well with the experiments for strong chemisorption. On the other hand, the deviation from the experimentally determined adsorption energy increases for weak bonding. For Ag and Au, the RPBE even predicts CO adsorption to be endothermic.
Furthermore, the RPBE functional removes the degeneracy of the adsorption sites for Co(0001) and Ag(111) surface and on-top sites become clearly favoured.
The favoured sites for CO on Rh, Pt, and Cu still contradict the experimental findings. However, in all these cases the corrugation of the potential energy surface is very small compared to metals where CO satisfies the experimental site preference (e.g. Ir and Pd \cite{pss:Over:58}). This also indicates that it is harder for the CO molecule to find the optimal adsorption site on these metals.

To make sure that this large change in E$_{\rm ads}$ is not the effect of the E$_{\rm cut}$ (larger basis set), but of the different functional, we have recalculated all systems with a harder PW91 pseudopotential which also requires an energy cut-off of E$_{\rm cut}$~=~700~eV. Besides a small reduction of the adsorption energies (weaker bonding) by about 50~meV the results are identical to those obtained for E$_{\rm cut}$~=~450~eV.

Summarizing, harder pseudopotentials for C and O lead to a slight decrease of the adsorption energies. The RPBE exchange-correlation functional reduces the adsorption energies by $\sim$0.3 eV and improves the differences in adsorption energies for different adsorption sites. However, the site preference remains the same as for the PW91 functional.

\section{Discussion}
\label{Discussion}

In this section, we discuss trends which can be deduced from our results for CO adsorption on closed-packed transition and noble metal surfaces.

\subsection{Correlations between adsorbate geometry, frequency and adsorption energy}

A strong correlation exists between the C-O bond length and the stretching frequency of the CO molecule $\nu_{\rm C-O}$. During adsorption the 2$\pi^{\star}$ orbital is partially populated and the C-O bond is weakened. This filling increases from the right to the left in the Periodic Table and with increasing coordination of the adsorption site and hence the CO bond length increases. A weaker and therefore longer C-O bond implies stronger CO-metal bonding and results in lower C-O stretching frequencies. The dependence of d$_{\rm C-O}$ on the coordination is smallest for the 3d and largest for 4d metals as can be seen in Fig. \ref{C-O bond length}. The trend goes as follows: the higher the reactivity of the surface, the stronger the bonding and population of 2$\pi^{\star}$-derived orbitals, resulting in a more elongated C-O bond length and a lower stretching frequency.

\begin{figure}[phtb]
\centerline{\psfig{figure=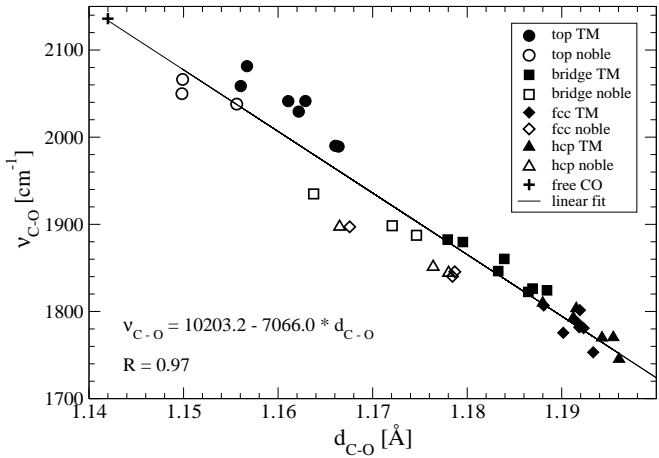,width=9.0cm,clip=true,angle=-90}}
\nopagebreak \caption{\label{figure freq_ap.geom_CO} Correlation
between the calculated CO bond length (d$_{\rm C-O}$) and the C-O stretching frequency ($\nu_{\rm C-O}$) for all high-symmetry adsorption sites.}
\end{figure}

In Fig. \ref{figure freq_ap.geom_CO} the CO stretching frequency
is plotted versus the molecular bond length exhibiting a linear
relationship between these two properties, which is almost
independent of the coordination of the CO molecule. The linear
regression fit is characterized by a correlation coefficient of
R~=~0.97.

\begin{figure}[phtb]
\centerline{\psfig{figure=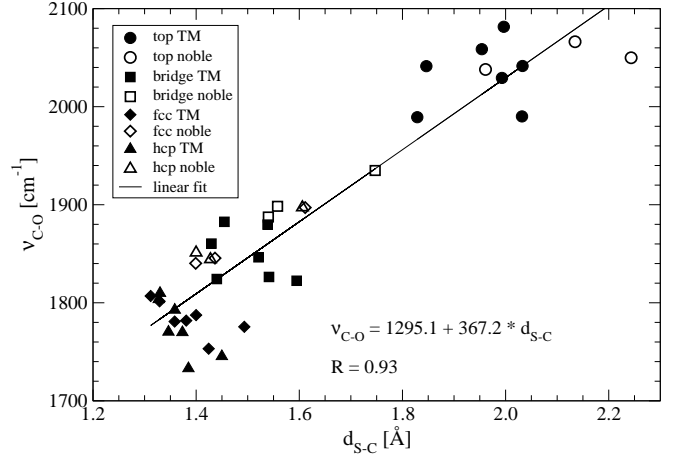,width=9.0cm,clip=true,angle=-90}}
\nopagebreak \caption{\label{figure freq_ap.geom_SC} Correlation
between the calculated S-C distance (d$_{\rm S-C}$) and the C-O stretching frequency ($\nu_{\rm C-O}$) for all high-symmetry adsorption sites. }
\end{figure}

As expected, pronounced stretching of the CO-bond (reflecting a stronger adsorbate-substrate interaction) is accompanied by a decrease of the adsorption height.
In Fig. \ref{figure freq_ap.geom_SC} we present this linear relationship between the vibrational frequency $\nu_{\rm CO}$ and the surface-carbon distance. Remarkably, a trend established by the noble metals alone would be characterized by a much bigger correlation coefficient of R~=~0.98 (fit is not shown) compared to the correlation coefficient of R~=~0.93 for the complete dataset.

Although we cannot directly relate the height of the molecule on
the surface to the adsorption energy it serves as a rough estimate
of the metal-adsorbate interaction. As a consequence of these two
linear relationships (d$_{\rm C-O}$ to $\nu_{\rm CO}$ and d$_{\rm
S-C}$ to d$_{\rm C-O}$) a similar correlation exists between the
d$_{\rm S-C}$ and d$_{\rm C-O}$  (Fig. \ref{figure
geom_SC.ap_geom_CO}). Again, we have divided our dataset into two
groups: noble and transition metals and we provide separately a
linear fit for each group of out set. Again, the correlation
coefficient is higher for noble (R$_{\rm noble}$~=~0.98) than for
transition metals (R$_{\rm TM}$~=~0.93).

Above mentioned linear relationships have an important consequence: if we know one of four measured properties ($\nu_{C-O}$, d$_{C-O}$, d$_{S-C}$, or d$_{M-C}$) we can easily estimate all other linearly related properties.

\begin{figure}[phtb]
\centerline{\psfig{figure=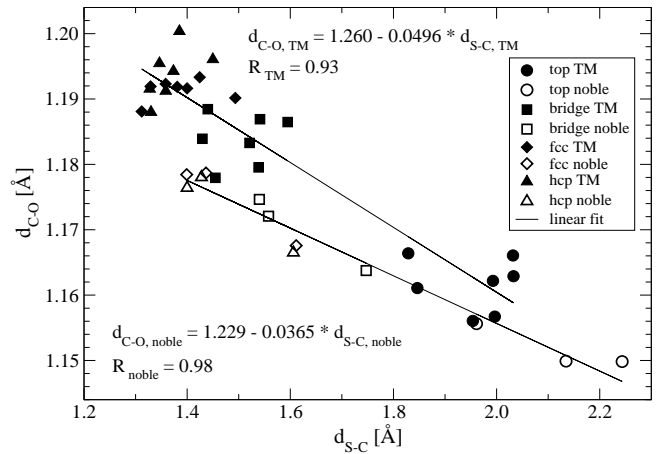,width=9.0cm,clip=true,angle=-90}}
\nopagebreak \caption{\label{figure geom_SC.ap_geom_CO} Correlation
between the calculated CO bond length (d$_{\rm C-O}$) and the S-C distance (d$_{\rm S-C}$) for all high-symmetry on-surface sites. }
\end{figure}

\subsection{Correlation between the electronic properties of the substrate and adsorption energy}

As already mentioned above, one of the most widely discussed
models for molecular adsorption on transition metal surfaces is
the d-band model of Hammer and N{\o }rskov
\cite{rpp:Norskov:53,ac:Hammer:45} relating the chemisorption
energy to the position of the metallic d-band relative to the
molecular orbitals of the adsorbate and to the coupling matrix
elements and overlap integrals between the metal d-states and
adsorbate states. In a simplified form of the argument, the
adsorption energy is related to the d-band position only. This
simplified d-band model has recently been examined by Lu et al.
\cite{jpca:Lu:106} for the adsorption of CO, H$_{\rm 2}$ and
ethylene. This analysis led the authors to a critique of the
d-band model and of DFT calculations of adsorption energies in
general. For CO it was pointed out that the reduction of the
adsorption energies with increasing distance of the d-band center
from the Fermi energy follows qualitatively the predicted trend,
but quantitatively the theoretically predicted dependence is much
stronger than what is observed experimentally.

\begin{figure}[phtb]
\centerline{\psfig{figure=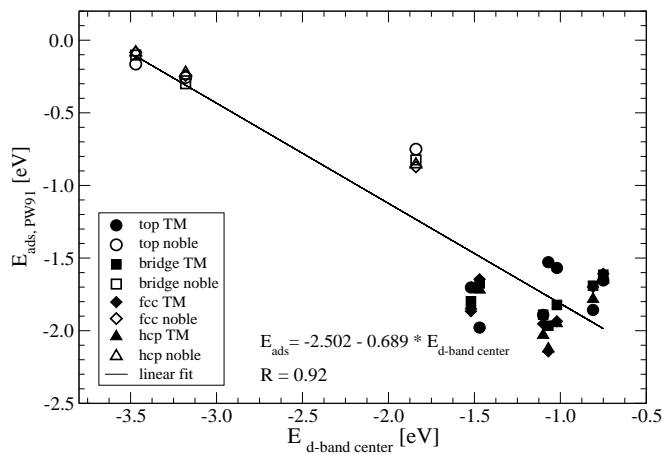,width=9.0cm,clip=true,angle=-90}}
\nopagebreak \caption{\label{figure dc_ads} Correlation
between the calculated position of the d-band center (where at least 9.5 electrons are accumulated, see Fig. \ref{figure dc clean}) for all high-symmetry on-surface sites. Empty symbols are for noble metals and full symbols are for transition metals.}
\end{figure}

 In Fig. \ref{figure dc_ads} we present calculated CO adsorption energies (PW91) with
respect to the d-band centers taken
from Fig. \ref{figure dc clean}. We can clearly establish a linear
relationship between position of the d-band center and the adsorption energy
for CO adsorption on noble metals. On the other hand, the
values for the transition metals show a large scatter,
hence what really determines the trend are the noble metals. It does
not mean that there is no relation between the adsorption energy
and the d-band center, but it must be corrected for the variation of the coupling matrix elements between different metals, as in the original argument of Hammer and N{\o }rskov \cite{ac:Hammer:45}.

\subsection{Site preference - limitation of DFT}

As a by-product of this study the performance of the PW91 functional with
different E$_{\rm cut}$ (harder potential) and the RPBE functional (well suited for CO and NO adsorption) could be tested on a large number of systems.

A calculation with harder pseudopotential for the same exchange-correlation functional (PW91) leads to small changes in the heat of adsorption which is closer to the experimental value for TMs, but further for noble metals. The site preference does not depend on the basis set for PW91 exchange-correlation functional.

The most important difference between the PW91 and RPBE functional
with more accurate pseudopotential (E$_{\rm cut}$ = 700 eV) is in
the absolute value of the of the adsorption energy $\Delta E_{\rm
ads}\sim$0.3~eV. Moreover, the difference in the adsorption energy
between previously almost degenerate top and hollow adsorption
sites is increased (Co, Ag, Au). However, all these changes do not
correct the prediction of a wrong adsorption site for Cu, Rh and
Pt.

Hence the question ``Why does DFT fail in predicting the right adsorption site as well as energy ?'' arises.
In the past, it has been proposed that a relativistic treatment of transition metals could solve the problem \cite{prb:Geschke:64,jcp:Grinberg:117,jcp:Olsen:119}. However, the results presented to support this suggestion are not really convincing, being based on rather thin slabs. Also we note that the work-function of the transition metals is well described and the electronic structure calculations are in good agreement with spectroscopic results.
 A more promising approach starts from the observation that 2$\pi^{\star}$-backdonation is the most important contribution to the adsorbate-substrate bond. It is well known that DFT calculations tend to underestimate the HOMO-LUMO gap in CO, facilitating back-donation. Very recently, Kresse et al. \cite{prb:Kresse:68} have used an LDA+U approach, leading to an increased HOMO-LUMO splitting due to an additional on-site Coulomb repulsion. It was shown that with an increasing HOMO-LUMO gap on Pt(111) 2$\pi^{\star}$ backdonation favouring hollow-adsorption is reduced and site-preference switches to the correct top-site.

Another way to improve the HOMO-LUMO gap is to use hybrid functionals mixing DFT-exchange with exact (Hartree-Fock) exchange: for gas-phase CO the PW91 functional predicts a  HOMO-LUMO gap of 6.8 eV, whereas the B3LYP hybrid functionals predicts a splitting of 10.1 eV. However, B3LYP calculations can at the moment only be performed using localized basis sets and small clusters. Very recently Gil et al. \cite{ss:Gil:530} have performed a series of cluster-calculations for CO on Pt(111) using both PW91 and B3LYP functionals and slab calculations using PW91. The conclusion is that the B3LYP functional reduces the overestimation of the stability of the hollow site compared to PW91 for a given cluster size. However, the results also show a pronounced cluster-size dependence: for small clusters the preference for the hollow-site is more pronounced than for larger clusters (the largest cluster considered in this study is Pt$_{\rm 19}$). For a comparison of cluster and slab calculations it is concluded that a slab calculation, if performed with a B3LYP functional should favour top adsorption, in agreement with experiment.

Current work is devoted to the extension of the LDA+U approach to other metals and to the investigation of the influence of meta-GGA functionals (incorporating a dependence on the kinetic energy density).

\section{Conclusions}
\label{Conclusion}

The adsorption behavior of the CO molecule on the
surfaces of closed-packed late transition metals, namely Co, Ni,
Cu, Ru, Rh, Pd, Pd, Ag, Ir, Pt, Au has been analyzed using ab-initio
DFT methods. To rationalize the trends in the CO adsorption we performed
investigations of the geometric structure (metal surface-C and C-O bond
length), the electronic structure (projected DOS, charge flow analysis) and the vibrational
properties ($\nu_{\rm C-O}$ and $\nu_{\rm M-CO}$). In addition, we examined the influence
of different GGAs (PW91, RPBE) on the CO site preference and adsorption energies.

We demonstrate that current DFT slightly overestimates substrate lattice
constants (1 to 2\%) and underestimates the work-function (by about 0.3 eV).
We compared our calculated values  for the geometric structure of the adsorbate-substrate complex with experimental data and we find them in very nice agreement. Furthermore, the C-O and M-CO
stretching frequencies agree well with experiments.
We illustrate the detailed interaction picture of the CO molecules for
different adsorption sites on TM surfaces by means of density of states projected onto molecular CO orbitals and change of the charge density due to adsorption.

Present density functionals overestimate
the adsorption energy for sites with high metal coordination compared to
sites with low metal coordination. Moreover, the underestimated energy difference between
HOMO ($5\sigma$) and LUMO ($2\pi^{\star}$) orbitals contributes to the CO over-binding.
DFT fails in predicting the site preference
for CO on the Cu (by $\eqslantgtr$ 30~meV), Rh (by $\eqslantgtr$ 30~meV), and Pt
(by $\eqslantgtr$ 90~meV) close-packed metal surfaces at 0.25~ML coverage.

Different GGA functionals reduce these discrepancies,
but neither harder potentials for CO, nor different GGA functionals correct the results for
CO adsorption on these substrates.
 One should also point out that the adsorption energies
 calculated using RPBE exchange-correlation functional give results which are much closer
 to the experimental values for TM surfaces, but predict CO adsorption to be endothermic on the noble metals (Ag and Au).

Recent results have shown that a correct site-preference can be achieved by correcting the HOMO-LUMO gap using techniques beyond DFT. This is after all not entirely surprising, since the analysis demonstrates the importance of the 2$\pi^{\star}$-backdonation effects, i.e. electron transfer to excited states of the free molecule.

%\bibliographystyle{plain}
%\bibliography{literature}

\begin{thebibliography}{97}
\expandafter\ifx\csname natexlab\endcsname\relax\def\natexlab#1{#1}\fi
\expandafter\ifx\csname bibnamefont\endcsname\relax
  \def\bibnamefont#1{#1}\fi
\expandafter\ifx\csname bibfnamefont\endcsname\relax
  \def\bibfnamefont#1{#1}\fi
\expandafter\ifx\csname citenamefont\endcsname\relax
  \def\citenamefont#1{#1}\fi
\expandafter\ifx\csname url\endcsname\relax
  \def\url#1{\texttt{#1}}\fi
\expandafter\ifx\csname urlprefix\endcsname\relax\def\urlprefix{URL }\fi
\providecommand{\bibinfo}[2]{#2}
\providecommand{\eprint}[2][]{\url{#2}}

\bibitem[{\citenamefont{Blyholder}(1964)}]{jpc:Blyholder:68}
\bibinfo{author}{\bibfnamefont{G.}~\bibnamefont{Blyholder}},
  \bibinfo{journal}{J. Phys. Chem.} \textbf{\bibinfo{volume}{68}},
  \bibinfo{pages}{2772} (\bibinfo{year}{1964}).

\bibitem[{\citenamefont{Bagus et~al.}(1983)\citenamefont{Bagus, Nelin, and
  Bauschlicher}}]{prb:Bagus:28}
\bibinfo{author}{\bibfnamefont{P.~S.} \bibnamefont{Bagus}},
  \bibinfo{author}{\bibfnamefont{C.~J.} \bibnamefont{Nelin}}, \bibnamefont{and}
  \bibinfo{author}{\bibfnamefont{C.~W.} \bibnamefont{Bauschlicher}},
  \bibinfo{journal}{Phys. Rev. B} \textbf{\bibinfo{volume}{28}},
  \bibinfo{pages}{5423} (\bibinfo{year}{1983}).

\bibitem[{\citenamefont{Over}(2001)}]{pss:Over:58}
\bibinfo{author}{\bibfnamefont{H.}~\bibnamefont{Over}}, \bibinfo{journal}{Prog.
  Surf. Sci.} \textbf{\bibinfo{volume}{58}}, \bibinfo{pages}{249}
  (\bibinfo{year}{2001}).

\bibitem[{\citenamefont{Sung and Hoffmann}(1985)}]{jacs:Sung:107}
\bibinfo{author}{\bibfnamefont{S.}~\bibnamefont{Sung}} \bibnamefont{and}
  \bibinfo{author}{\bibfnamefont{R.}~\bibnamefont{Hoffmann}},
  \bibinfo{journal}{J. Am. Chem. Soc.} \textbf{\bibinfo{volume}{107}},
  \bibinfo{pages}{578} (\bibinfo{year}{1985}).

\bibitem[{\citenamefont{Feibelman et~al.}(2001)\citenamefont{Feibelman, Hammer,
  N{\o }^{^{}}rskov, Wagner, Scheffler, and Stumpf}}]{jpcb:Feibelman:105}
\bibinfo{author}{\bibfnamefont{P.}~\bibnamefont{Feibelman}},
  \bibinfo{author}{\bibfnamefont{B.}~\bibnamefont{Hammer}},
  \bibinfo{author}{\bibfnamefont{J.}~\bibnamefont{N{\o }rskov}},
  \bibinfo{author}{\bibfnamefont{F.}~\bibnamefont{Wagner}},
  \bibinfo{author}{\bibfnamefont{M.}~\bibnamefont{Scheffler}},
  \bibnamefont{and} \bibinfo{author}{\bibfnamefont{R.}~\bibnamefont{Stumpf}},
  \bibinfo{journal}{J. Phys. Chem. B} \textbf{\bibinfo{volume}{108}},
  \bibinfo{pages}{4018} (\bibinfo{year}{2001}).


\bibitem[{\citenamefont{Ying et~al.}(1975)\citenamefont{Ying, Smith, and
  Kohn}}]{prb:Ying:11}
\bibinfo{author}{\bibfnamefont{S.~C.} \bibnamefont{Ying}},
  \bibinfo{author}{\bibfnamefont{J.~R.} \bibnamefont{Smith}}, \bibnamefont{and}
  \bibinfo{author}{\bibfnamefont{W.}~\bibnamefont{Kohn}},
  \bibinfo{journal}{Phys. Rev. B} \textbf{\bibinfo{volume}{11}},
  \bibinfo{pages}{1483} (\bibinfo{year}{1975}).


\bibitem[{\citenamefont{Andreoni and Varma}(1981)}]{prb:Andreoni:23}
\bibinfo{author}{\bibfnamefont{W.}~\bibnamefont{Andreoni}} \bibnamefont{and}
  \bibinfo{author}{\bibfnamefont{C.~M.} \bibnamefont{Varma}},
  \bibinfo{journal}{Phys. Rev. B} \textbf{\bibinfo{volume}{23}},
  \bibinfo{pages}{437} (\bibinfo{year}{1981}).

\bibitem[{\citenamefont{Hammer and N{\o }rskov}(2000)}]{ac:Hammer:45}
\bibinfo{author}{\bibfnamefont{B.}~\bibnamefont{Hammer}} \bibnamefont{and}
  \bibinfo{author}{\bibfnamefont{J.~K.} \bibnamefont{N{\o }rskov}},
  \bibinfo{journal}{Advances in Catalysis} \textbf{\bibinfo{volume}{45}},
  \bibinfo{pages}{71} (\bibinfo{year}{2000}).

\bibitem[{\citenamefont{N{\o }rskov}(1990)}]{rpp:Norskov:53}
\bibinfo{author}{\bibfnamefont{J.~K.} \bibnamefont{N{\o }rskov}},
  \bibinfo{journal}{Rep. Prog. Phys.} \textbf{\bibinfo{volume}{53}},
  \bibinfo{pages}{1253} (\bibinfo{year}{1990}).

\bibitem[{\citenamefont{Ishi et~al.}(1985)\citenamefont{Ishi, Ohno, and
  Viswanathan}}]{ss:Ishi:161}
\bibinfo{author}{\bibfnamefont{S.}~\bibnamefont{Ishi}},
  \bibinfo{author}{\bibfnamefont{Y.}~\bibnamefont{Ohno}}, \bibnamefont{and}
  \bibinfo{author}{\bibfnamefont{B.}~\bibnamefont{Viswanathan}},
  \bibinfo{journal}{Surf. Sci.} \textbf{\bibinfo{volume}{161}},
  \bibinfo{pages}{349} (\bibinfo{year}{1985}).

\bibitem[{\citenamefont{http://cms.mpi.univie.ac.at/vasp/}()}]{vasp}
\bibinfo{author}{\bibnamefont{http://cms.mpi.univie.ac.at/vasp/}}.

\bibitem[{\citenamefont{Kresse and Furthm\"uller}(1996)}]{prb:Kresse:54}
\bibinfo{author}{\bibfnamefont{G.}~\bibnamefont{Kresse}} \bibnamefont{and}
  \bibinfo{author}{\bibfnamefont{J.}~\bibnamefont{Furthm\"uller}},
  \bibinfo{journal}{Phys. Rev. B} \textbf{\bibinfo{volume}{54}},
  \bibinfo{pages}{11169} (\bibinfo{year}{1996}).

\bibitem[{\citenamefont{Bl\"ochl}(1994)}]{prb:Blochl:50}
\bibinfo{author}{\bibfnamefont{P.}~\bibnamefont{Bl\"ochl}},
  \bibinfo{journal}{Phys. Rev. B} \textbf{\bibinfo{volume}{50}},
  \bibinfo{pages}{17953} (\bibinfo{year}{1994}).

\bibitem[{\citenamefont{Kresse and Joubert}(1999)}]{prb:Kresse:59}
\bibinfo{author}{\bibfnamefont{G.}~\bibnamefont{Kresse}} \bibnamefont{and}
  \bibinfo{author}{\bibfnamefont{D.}~\bibnamefont{Joubert}},
  \bibinfo{journal}{Phys. Rev. B} \textbf{\bibinfo{volume}{59}},
  \bibinfo{pages}{1758} (\bibinfo{year}{1999}).

\bibitem[{\citenamefont{Perdew and Zunger}(1981)}]{prb:Perdew:23}
\bibinfo{author}{\bibfnamefont{J.~P.} \bibnamefont{Perdew}} \bibnamefont{and}
  \bibinfo{author}{\bibfnamefont{A.}~\bibnamefont{Zunger}},
  \bibinfo{journal}{Phys. Rev. B} \textbf{\bibinfo{volume}{23}},
  \bibinfo{pages}{5048} (\bibinfo{year}{1981}).

\bibitem[{\citenamefont{Perdew et~al.}(1992)\citenamefont{Perdew, Chevary,
  Vosko, Jackson, Pederson, Singh, and Fiolhais}}]{prb:Perdew:46}
\bibinfo{author}{\bibfnamefont{J.~P.} \bibnamefont{Perdew}},
  \bibinfo{author}{\bibfnamefont{J.~A.} \bibnamefont{Chevary}},
  \bibinfo{author}{\bibfnamefont{S.~H.} \bibnamefont{Vosko}},
  \bibinfo{author}{\bibfnamefont{K.~A.} \bibnamefont{Jackson}},
  \bibinfo{author}{\bibfnamefont{M.~R.} \bibnamefont{Pederson}},
  \bibinfo{author}{\bibfnamefont{D.~J.} \bibnamefont{Singh}}, \bibnamefont{and}
  \bibinfo{author}{\bibfnamefont{C.}~\bibnamefont{Fiolhais}},
  \bibinfo{journal}{Phys. Rev. B} \textbf{\bibinfo{volume}{46}},
  \bibinfo{pages}{6671} (\bibinfo{year}{1992}).

\bibitem[{\citenamefont{Hammer et~al.}(1999)\citenamefont{Hammer, Hansen, and
  N{\o }rskov}}]{prb:Hammer:59}
\bibinfo{author}{\bibfnamefont{B.}~\bibnamefont{Hammer}},
  \bibinfo{author}{\bibfnamefont{L.~B.} \bibnamefont{Hansen}},
  \bibnamefont{and} \bibinfo{author}{\bibfnamefont{J.~K.} \bibnamefont{N{\o
  }rskov}}, \bibinfo{journal}{Phys. Rev. B} \textbf{\bibinfo{volume}{59}},
  \bibinfo{pages}{7413} (\bibinfo{year}{1999}).

\bibitem[{\citenamefont{Mantz et~al.}(1971)\citenamefont{Mantz, Watson, Rao,
  Albritton, Schmeltekope, and Zare}}]{jms:Mantz:39}
\bibinfo{author}{\bibfnamefont{A.~W.} \bibnamefont{Mantz}},
  \bibinfo{author}{\bibfnamefont{J.~K.~G.} \bibnamefont{Watson}},
  \bibinfo{author}{\bibfnamefont{K.~N.} \bibnamefont{Rao}},
  \bibinfo{author}{\bibfnamefont{D.~L.} \bibnamefont{Albritton}},
  \bibinfo{author}{\bibfnamefont{A.~L.} \bibnamefont{Schmeltekope}},
  \bibnamefont{and} \bibinfo{author}{\bibfnamefont{R.~N.} \bibnamefont{Zare}},
  \bibinfo{journal}{J. Mol. Spectrosc.} \textbf{\bibinfo{volume}{39}},
  \bibinfo{pages}{180} (\bibinfo{year}{1971}).

\bibitem[{\citenamefont{Cox et~al.}(1984)\citenamefont{Cox, Wangman, and
  Medvedev}}]{book:Cox:1984}
\bibinfo{author}{\bibfnamefont{J.~D.} \bibnamefont{Cox}},
  \bibinfo{author}{\bibfnamefont{D.~D.} \bibnamefont{Wangman}},
  \bibnamefont{and} \bibinfo{author}{\bibfnamefont{V.~A.}
  \bibnamefont{Medvedev}}, \emph{\bibinfo{title}{CODATA Key Values for
  Thermodynamics}} (\bibinfo{publisher}{Hemisphere Publishing Corp.},
  \bibinfo{year}{1984}).

\bibitem[{\citenamefont{Landolt-B\"ornstein}(1971)}]{springer:Landolt:1971}
\bibinfo{author}{\bibnamefont{Landolt-B\"ornstein}},
  \emph{\bibinfo{title}{Strukturdaten der Elemente und intermetallischer
  Phasen}}, \emph{\bibinfo{series}{Neue Serie}}, vol. \bibinfo{volume}{III}
  (\bibinfo{publisher}{Springer}, \bibinfo{year}{1971}).

\bibitem[{\citenamefont{Wu and Metiu}(2000)}]{jcp:Wu:113}
\bibinfo{author}{\bibfnamefont{M.~W.} \bibnamefont{Wu}} \bibnamefont{and}
  \bibinfo{author}{\bibfnamefont{H.}~\bibnamefont{Metiu}}, \bibinfo{journal}{J.
  Chem. Phys.} \textbf{\bibinfo{volume}{113}}, \bibinfo{pages}{1177}
  (\bibinfo{year}{2000}).

\bibitem[{\citenamefont{Mavrikakis et~al.}(1998)\citenamefont{Mavrikakis,
  Hammer, and N{\o }rskov}}]{prl:Mavrikakis:81}
\bibinfo{author}{\bibfnamefont{M.}~\bibnamefont{Mavrikakis}},
  \bibinfo{author}{\bibfnamefont{B.}~\bibnamefont{Hammer}}, \bibnamefont{and}
  \bibinfo{author}{\bibfnamefont{J.~K.} \bibnamefont{N{\o }rskov}},
  \bibinfo{journal}{Phys. Rev. Lett.} \textbf{\bibinfo{volume}{81}},
  \bibinfo{pages}{2819} (\bibinfo{year}{1998}).

\bibitem[{\citenamefont{Michaelson}(1977)}]{jap:Michaelson:48}
\bibinfo{author}{\bibfnamefont{H.~B.} \bibnamefont{Michaelson}},
  \bibinfo{journal}{J. Appl. Phys.} \textbf{\bibinfo{volume}{48}},
  \bibinfo{pages}{4729} (\bibinfo{year}{1977}).

\bibitem[{\citenamefont{Himpsel et~al.}(1982)\citenamefont{Himpsel, Christmann,
  Heimann, Eastmann, and Feibelmann}}]{ss:Himpsel:115}
\bibinfo{author}{\bibfnamefont{F.~J.} \bibnamefont{Himpsel}},
  \bibinfo{author}{\bibfnamefont{K.}~\bibnamefont{Christmann}},
  \bibinfo{author}{\bibfnamefont{P.}~\bibnamefont{Heimann}},
  \bibinfo{author}{\bibfnamefont{D.~E.} \bibnamefont{Eastmann}},
  \bibnamefont{and} \bibinfo{author}{\bibfnamefont{P.~J.}
  \bibnamefont{Feibelmann}}, \bibinfo{journal}{Surf. Sci.}
  \textbf{\bibinfo{volume}{115}}, \bibinfo{pages}{L159} (\bibinfo{year}{1982}).

\bibitem[{\citenamefont{B{\"o}ttcher and Niehus}(1999)}]{prb:Bottcher:60}
\bibinfo{author}{\bibfnamefont{A.}~\bibnamefont{B{\"o}ttcher}}
  \bibnamefont{and} \bibinfo{author}{\bibfnamefont{H.}~\bibnamefont{Niehus}},
  \bibinfo{journal}{Phys. Rev. B} \textbf{\bibinfo{volume}{60}},
  \bibinfo{pages}{14396} (\bibinfo{year}{1999}).

\bibitem[{\citenamefont{Hendrickx}(1988)}]{phd:Hendrickx}
\bibinfo{author}{\bibfnamefont{H.~A. C.~M.} \bibnamefont{Hendrickx}},
  \bibinfo{journal}{PhD Thesis, Rijksuniversiteit Leiden, The Netherlands}
  (\bibinfo{year}{1988}).

\bibitem[{\citenamefont{Fischer et~al.}(1993)\citenamefont{Fischer, Schuppler,
  Fischer, Fauster, and Steineman}}]{prl:Fischer:70}
\bibinfo{author}{\bibfnamefont{R.}~\bibnamefont{Fischer}},
  \bibinfo{author}{\bibfnamefont{S.}~\bibnamefont{Schuppler}},
  \bibinfo{author}{\bibfnamefont{N.}~\bibnamefont{Fischer}},
  \bibinfo{author}{\bibfnamefont{T.}~\bibnamefont{Fauster}}, \bibnamefont{and}
  \bibinfo{author}{\bibfnamefont{W.}~\bibnamefont{Steineman}},
  \bibinfo{journal}{Phys. Rev. Lett.} \textbf{\bibinfo{volume}{70}},
  \bibinfo{pages}{654} (\bibinfo{year}{1993}).

\bibitem[{\citenamefont{Nieuwenhuys}(1973)}]{ss:Nieuwenhuys:34}
\bibinfo{author}{\bibfnamefont{B.~E.} \bibnamefont{Nieuwenhuys}},
  \bibinfo{journal}{Surf. Sci.} \textbf{\bibinfo{volume}{34}},
  \bibinfo{pages}{317} (\bibinfo{year}{1973}).

\bibitem[{\citenamefont{Hammer and N{\o }rskov}(1995)}]{ss:Hammer:343}
\bibinfo{author}{\bibfnamefont{B.}~\bibnamefont{Hammer}} \bibnamefont{and}
  \bibinfo{author}{\bibfnamefont{J.~K.} \bibnamefont{N{\o }rskov}},
  \bibinfo{journal}{Surf. Sci.} \textbf{\bibinfo{volume}{343}},
  \bibinfo{pages}{211} (\bibinfo{year}{1995}).

\bibitem[{\citenamefont{Hammer et~al.}(1996)\citenamefont{Hammer, Morikawa, and
  N{\o }rskov}}]{prl:Hammer:76}
\bibinfo{author}{\bibfnamefont{B.}~\bibnamefont{Hammer}},
  \bibinfo{author}{\bibfnamefont{Y.}~\bibnamefont{Morikawa}}, \bibnamefont{and}
  \bibinfo{author}{\bibfnamefont{J.~K.} \bibnamefont{N{\o }rskov}},
  \bibinfo{journal}{Phys. Rev. Lett.} \textbf{\bibinfo{volume}{76}},
  \bibinfo{pages}{2141} (\bibinfo{year}{1996}).

\bibitem[{\citenamefont{Somorjai}(1979)}]{ss:Somorjai:89}
\bibinfo{author}{\bibfnamefont{G.}~\bibnamefont{Somorjai}},
  \bibinfo{journal}{Surf. Sci.} \textbf{\bibinfo{volume}{89}},
  \bibinfo{pages}{496} (\bibinfo{year}{1979}).

\bibitem[{\citenamefont{Papp}(1983)}]{ss:Papp:129}
\bibinfo{author}{\bibfnamefont{H.}~\bibnamefont{Papp}}, \bibinfo{journal}{Surf.
  Sci.} \textbf{\bibinfo{volume}{129}}, \bibinfo{pages}{205}
  (\bibinfo{year}{1983}).

\bibitem[{\citenamefont{Lahtinen et~al.}(1988)\citenamefont{Lahtinen, Vaari,
  and Kauraala}}]{ss:Lahtinen:418}
\bibinfo{author}{\bibfnamefont{J.}~\bibnamefont{Lahtinen}},
  \bibinfo{author}{\bibfnamefont{J.}~\bibnamefont{Vaari}}, \bibnamefont{and}
  \bibinfo{author}{\bibfnamefont{K.}~\bibnamefont{Kauraala}},
  \bibinfo{journal}{Surf. Sci.} \textbf{\bibinfo{volume}{418}},
  \bibinfo{pages}{502} (\bibinfo{year}{1988}).

\bibitem[{\citenamefont{Stuckless et~al.}(1993)\citenamefont{Stuckless,
  Al-Sarraf, Wartnaby, , and King}}]{jcp:Stuckless:99}
\bibinfo{author}{\bibfnamefont{J.~T.} \bibnamefont{Stuckless}},
  \bibinfo{author}{\bibfnamefont{N.}~\bibnamefont{Al-Sarraf}},
  \bibinfo{author}{\bibfnamefont{C.}~\bibnamefont{Wartnaby}}, ,
  \bibnamefont{and} \bibinfo{author}{\bibfnamefont{D.~A.} \bibnamefont{King}},
  \bibinfo{journal}{J. Chem. Phys.} \textbf{\bibinfo{volume}{203}},
  \bibinfo{pages}{2202} (\bibinfo{year}{1993}).

\bibitem[{\citenamefont{Froitzheim and Koehler}(1987)}]{ss:Froitzheim:188}
\bibinfo{author}{\bibfnamefont{H.}~\bibnamefont{Froitzheim}} \bibnamefont{and}
  \bibinfo{author}{\bibfnamefont{U.}~\bibnamefont{Koehler}},
  \bibinfo{journal}{Surf. Sci.} \textbf{\bibinfo{volume}{188}},
  \bibinfo{pages}{70} (\bibinfo{year}{1987}).

\bibitem[{\citenamefont{Miller et~al.}(1987)\citenamefont{Miller, Siddiqui,
  Gates, Russell{ Jr.}, Yates{ Jr.}, Tully, and Cardillo}}]{jcp:Miller:87}
\bibinfo{author}{\bibfnamefont{J.~B.} \bibnamefont{Miller}},
  \bibinfo{author}{\bibfnamefont{H.~R.} \bibnamefont{Siddiqui}},
  \bibinfo{author}{\bibfnamefont{S.~M.} \bibnamefont{Gates}},
  \bibinfo{author}{\bibfnamefont{J.~N.} \bibnamefont{Russell{ Jr.}}},
  \bibinfo{author}{\bibfnamefont{J.~T.} \bibnamefont{Yates{ Jr.}}},
  \bibinfo{author}{\bibfnamefont{J.~C.} \bibnamefont{Tully}}, \bibnamefont{and}
  \bibinfo{author}{\bibfnamefont{M.~J.} \bibnamefont{Cardillo}},
  \bibinfo{journal}{J. Chem. Phys.} \textbf{\bibinfo{volume}{87}},
  \bibinfo{pages}{6725} (\bibinfo{year}{1987}).

\bibitem[{\citenamefont{Vollmer et~al.}(2001)\citenamefont{Vollmer, Witte, and
  Woell}}]{cl:Vollmer:77}
\bibinfo{author}{\bibfnamefont{S.}~\bibnamefont{Vollmer}},
  \bibinfo{author}{\bibfnamefont{G.}~\bibnamefont{Witte}}, \bibnamefont{and}
  \bibinfo{author}{\bibfnamefont{C.}~\bibnamefont{Woell}},
  \bibinfo{journal}{Catal. Lett.} \textbf{\bibinfo{volume}{77}},
  \bibinfo{pages}{97} (\bibinfo{year}{2001}).

\bibitem[{\citenamefont{Kirstein et~al.}(1986)\citenamefont{Kirstein,
  Kr\"ueger, and Thieme}}]{ss:Kirstein:176}
\bibinfo{author}{\bibfnamefont{W.}~\bibnamefont{Kirstein}},
  \bibinfo{author}{\bibfnamefont{B.}~\bibnamefont{Kr\"ueger}},
  \bibnamefont{and} \bibinfo{author}{\bibfnamefont{F.}~\bibnamefont{Thieme}},
  \bibinfo{journal}{Surf. Sci.} \textbf{\bibinfo{volume}{176}},
  \bibinfo{pages}{505} (\bibinfo{year}{1986}).

\bibitem[{\citenamefont{Kessler and Thieme}(1977)}]{ss:Kessler:67}
\bibinfo{author}{\bibfnamefont{J.}~\bibnamefont{Kessler}} \bibnamefont{and}
  \bibinfo{author}{\bibfnamefont{F.}~\bibnamefont{Thieme}},
  \bibinfo{journal}{Surf. Sci.} \textbf{\bibinfo{volume}{67}},
  \bibinfo{pages}{405} (\bibinfo{year}{1977}).

\bibitem[{\citenamefont{Pfn\"ur and Menzel}(1983)}]{jcp:Pfnuer:79}
\bibinfo{author}{\bibfnamefont{H.}~\bibnamefont{Pfn\"ur}} \bibnamefont{and}
  \bibinfo{author}{\bibfnamefont{D.}~\bibnamefont{Menzel}},
  \bibinfo{journal}{J. Chem. Phys.} \textbf{\bibinfo{volume}{79}},
  \bibinfo{pages}{4613} (\bibinfo{year}{1983}).

\bibitem[{\citenamefont{M.~Smedh et~al.}(2001)\citenamefont{M.~Smedh, Borg,
  Nyholm, and Andersen}}]{ss:Smedh:491:115}
\bibinfo{author}{\bibfnamefont{A.~B.} \bibnamefont{M.~Smedh}},
  \bibinfo{author}{\bibfnamefont{M.}~\bibnamefont{Borg}},
  \bibinfo{author}{\bibfnamefont{R.}~\bibnamefont{Nyholm}}, \bibnamefont{and}
  \bibinfo{author}{\bibfnamefont{J.~N.} \bibnamefont{Andersen}},
  \bibinfo{journal}{Surf. Sci.} \textbf{\bibinfo{volume}{491}},
  \bibinfo{pages}{115} (\bibinfo{year}{2001}).

\bibitem[{\citenamefont{Thiel et~al.}(1979)\citenamefont{Thiel, Wiliams, and
  Yates}}]{ss:Thiel:84}
\bibinfo{author}{\bibfnamefont{P.~A.} \bibnamefont{Thiel}},
  \bibinfo{author}{\bibfnamefont{E.~D.} \bibnamefont{Wiliams}},
  \bibnamefont{and} \bibinfo{author}{\bibfnamefont{J.~T.} \bibnamefont{Yates}},
  \bibinfo{journal}{Surf. Sci.} \textbf{\bibinfo{volume}{84}},
  \bibinfo{pages}{54} (\bibinfo{year}{1979}).

\bibitem[{\citenamefont{Wei et~al.}(1997)\citenamefont{Wei, Skelton, and
  Kevan}}]{ss:Wei:381}
\bibinfo{author}{\bibfnamefont{D.~H.} \bibnamefont{Wei}},
  \bibinfo{author}{\bibfnamefont{D.~C.} \bibnamefont{Skelton}},
  \bibnamefont{and} \bibinfo{author}{\bibfnamefont{S.~D.} \bibnamefont{Kevan}},
  \bibinfo{journal}{Surf. Sci.} \textbf{\bibinfo{volume}{381}},
  \bibinfo{pages}{49} (\bibinfo{year}{1997}).

\bibitem[{\citenamefont{Perlinz et~al.}(1991)\citenamefont{Perlinz, Curtiss,
  and Sibener}}]{jcp:Perlinz:95}
\bibinfo{author}{\bibfnamefont{K.~A.} \bibnamefont{Perlinz}},
  \bibinfo{author}{\bibfnamefont{T.~J.} \bibnamefont{Curtiss}},
  \bibnamefont{and} \bibinfo{author}{\bibfnamefont{S.~J.}
  \bibnamefont{Sibener}}, \bibinfo{journal}{J. Chem. Phys.}
  \textbf{\bibinfo{volume}{95}}, \bibinfo{pages}{6972} (\bibinfo{year}{1991}).

\bibitem[{\citenamefont{Szanyi et~al.}(1993)\citenamefont{Szanyi, Kuhn, and
  Goodman}}]{jvst:Szanyi:11}
\bibinfo{author}{\bibfnamefont{J.}~\bibnamefont{Szanyi}},
  \bibinfo{author}{\bibfnamefont{W.~K.} \bibnamefont{Kuhn}}, \bibnamefont{and}
  \bibinfo{author}{\bibfnamefont{D.~W.} \bibnamefont{Goodman}},
  \bibinfo{journal}{J. Vac. Sci. Technol. A} \textbf{\bibinfo{volume}{11}},
  \bibinfo{pages}{1969} (\bibinfo{year}{1993}).

\bibitem[{\citenamefont{Guo and Yates{ Jr.}}(1989)}]{jcp:Guo:90}
\bibinfo{author}{\bibfnamefont{X.}~\bibnamefont{Guo}} \bibnamefont{and}
  \bibinfo{author}{\bibfnamefont{J.}~\bibnamefont{Yates{ Jr.}}},
  \bibinfo{journal}{J. Chem. Phys.} \textbf{\bibinfo{volume}{90}},
  \bibinfo{pages}{6761} (\bibinfo{year}{1989}).

\bibitem[{\citenamefont{McElhiney et~al.}(1976)\citenamefont{McElhiney, Papp,
  and Pritchard}}]{ss:McElhiney:54}
\bibinfo{author}{\bibfnamefont{G.}~\bibnamefont{McElhiney}},
  \bibinfo{author}{\bibfnamefont{H.}~\bibnamefont{Papp}}, \bibnamefont{and}
  \bibinfo{author}{\bibfnamefont{J.}~\bibnamefont{Pritchard}},
  \bibinfo{journal}{Surf. Sci.} \textbf{\bibinfo{volume}{54}},
  \bibinfo{pages}{617} (\bibinfo{year}{1976}).

\bibitem[{\citenamefont{Sushchikh et~al.}(1997)\citenamefont{Sushchikh,
  Lauterbach, and Weinberg}}]{jvst:Sushchikh:15}
\bibinfo{author}{\bibfnamefont{M.}~\bibnamefont{Sushchikh}},
  \bibinfo{author}{\bibfnamefont{J.}~\bibnamefont{Lauterbach}},
  \bibnamefont{and} \bibinfo{author}{\bibfnamefont{W.~H.}
  \bibnamefont{Weinberg}}, \bibinfo{journal}{J. Vac. Sci. Technol. A}
  \textbf{\bibinfo{volume}{15}}, \bibinfo{pages}{1630} (\bibinfo{year}{1997}).

\bibitem[{\citenamefont{Comrie and Weinberg}(1976)}]{jcp:Comrie:64}
\bibinfo{author}{\bibfnamefont{C.~M.} \bibnamefont{Comrie}} \bibnamefont{and}
  \bibinfo{author}{\bibfnamefont{W.~H.} \bibnamefont{Weinberg}},
  \bibinfo{journal}{J. Chem. Phys.} \textbf{\bibinfo{volume}{64}},
  \bibinfo{pages}{250} (\bibinfo{year}{1976}).

\bibitem[{\citenamefont{Steininger et~al.}(1982)\citenamefont{Steininger,
  Lehwald, and Ibach}}]{ss:Steininger:123}
\bibinfo{author}{\bibfnamefont{H.}~\bibnamefont{Steininger}},
  \bibinfo{author}{\bibfnamefont{S.}~\bibnamefont{Lehwald}}, \bibnamefont{and}
  \bibinfo{author}{\bibfnamefont{H.}~\bibnamefont{Ibach}},
  \bibinfo{journal}{Surf. Sci.} \textbf{\bibinfo{volume}{123}},
  \bibinfo{pages}{264} (\bibinfo{year}{1982}).

\bibitem[{\citenamefont{Seebauer et~al.}(1987)\citenamefont{Seebauer, Kong, and
  Schmidt}}]{jvst:Seebauer:5}
\bibinfo{author}{\bibfnamefont{E.~G.} \bibnamefont{Seebauer}},
  \bibinfo{author}{\bibfnamefont{A.~C.~F.} \bibnamefont{Kong}},
  \bibnamefont{and} \bibinfo{author}{\bibfnamefont{L.~D.}
  \bibnamefont{Schmidt}}, \bibinfo{journal}{J. Vac. Sci. Technol.}
  \textbf{\bibinfo{volume}{5}}, \bibinfo{pages}{464} (\bibinfo{year}{1987}).

\bibitem[{\citenamefont{Ertl et~al.}(1977)\citenamefont{Ertl, Neumann, and
  Streit}}]{ss:Ertl:64}
\bibinfo{author}{\bibfnamefont{G.}~\bibnamefont{Ertl}},
  \bibinfo{author}{\bibfnamefont{M.}~\bibnamefont{Neumann}}, \bibnamefont{and}
  \bibinfo{author}{\bibfnamefont{K.~M.} \bibnamefont{Streit}},
  \bibinfo{journal}{Surf. Sci.} \textbf{\bibinfo{volume}{64}},
  \bibinfo{pages}{393} (\bibinfo{year}{1977}).

\bibitem[{\citenamefont{Elliott and Miller}(1984)}]{Elliott}
\bibinfo{author}{\bibfnamefont{G.~S.} \bibnamefont{Elliott}} \bibnamefont{and}
  \bibinfo{author}{\bibfnamefont{D.~R.} \bibnamefont{Miller}}, in
  \emph{\bibinfo{booktitle}{Proceedings of the 14. International Symposium on
  Rarefied Gas Dynamics}} (\bibinfo{address}{Tokyo}, \bibinfo{year}{1984}), pp.
  \bibinfo{pages}{565, and personal communication}.

\bibitem[{\citenamefont{Westerlund et~al.}(1988)\citenamefont{Westerlund,
  J\"onsson, and Andersson}}]{ss:Westerlund:199}
\bibinfo{author}{\bibfnamefont{L.}~\bibnamefont{Westerlund}},
  \bibinfo{author}{\bibfnamefont{L.}~\bibnamefont{J\"onsson}},
  \bibnamefont{and}
  \bibinfo{author}{\bibfnamefont{S.}~\bibnamefont{Andersson}},
  \bibinfo{journal}{Surf. Sci.} \textbf{\bibinfo{volume}{199}},
  \bibinfo{pages}{109} (\bibinfo{year}{1988}).

\bibitem[{\citenamefont{Lahtinen et~al.}(2000)\citenamefont{Lahtinen, Vaari,
  Kauraala, Soares, and {Van }Hove}}]{ss:Lahtinen:448}
\bibinfo{author}{\bibfnamefont{J.}~\bibnamefont{Lahtinen}},
  \bibinfo{author}{\bibfnamefont{J.}~\bibnamefont{Vaari}},
  \bibinfo{author}{\bibfnamefont{K.}~\bibnamefont{Kauraala}},
  \bibinfo{author}{\bibfnamefont{E.~A.} \bibnamefont{Soares}},
  \bibnamefont{and} \bibinfo{author}{\bibfnamefont{M.~A.} \bibnamefont{{Van
  }Hove}}, \bibinfo{journal}{Surf. Sci.} \textbf{\bibinfo{volume}{448}},
  \bibinfo{pages}{269} (\bibinfo{year}{2000}).

\bibitem[{\citenamefont{Eichler}(2003)}]{ss:Eichler:526}
\bibinfo{author}{\bibfnamefont{A.}~\bibnamefont{Eichler}},
  \bibinfo{journal}{Surf. Sci.} \textbf{\bibinfo{volume}{526}},
  \bibinfo{pages}{332} (\bibinfo{year}{2003}).

\bibitem[{\citenamefont{Moler et~al.}(1996)\citenamefont{Moler, Kellar, Huff,
  and Hussain}}]{prb:Moler:54}
\bibinfo{author}{\bibfnamefont{E.~J.} \bibnamefont{Moler}},
  \bibinfo{author}{\bibfnamefont{S.~A.} \bibnamefont{Kellar}},
  \bibinfo{author}{\bibfnamefont{W.~R.~A.} \bibnamefont{Huff}},
  \bibnamefont{and} \bibinfo{author}{\bibfnamefont{Z.}~\bibnamefont{Hussain}},
  \bibinfo{journal}{Phys. Rev. B} \textbf{\bibinfo{volume}{54}},
  \bibinfo{pages}{10862} (\bibinfo{year}{1996}).

\bibitem[{\citenamefont{Over et~al.}(1993)\citenamefont{Over, Moritz, and
  Ertl}}]{prl:Over:70}
\bibinfo{author}{\bibfnamefont{H.}~\bibnamefont{Over}},
  \bibinfo{author}{\bibfnamefont{W.}~\bibnamefont{Moritz}}, \bibnamefont{and}
  \bibinfo{author}{\bibfnamefont{G.}~\bibnamefont{Ertl}},
  \bibinfo{journal}{Phys. Rev. Lett.} \textbf{\bibinfo{volume}{70}},
  \bibinfo{pages}{315} (\bibinfo{year}{1993}).

\bibitem[{\citenamefont{Gierer et~al.}(1997)\citenamefont{Gierer, Barbieri,
  {Van }Hove, and Somorjai}}]{ss:Gierer:391}
\bibinfo{author}{\bibfnamefont{M.}~\bibnamefont{Gierer}},
  \bibinfo{author}{\bibfnamefont{A.}~\bibnamefont{Barbieri}},
  \bibinfo{author}{\bibfnamefont{M.~A.} \bibnamefont{{Van }Hove}},
  \bibnamefont{and} \bibinfo{author}{\bibfnamefont{G.~A.}
  \bibnamefont{Somorjai}}, \bibinfo{journal}{Surf. Sci.}
  \textbf{\bibinfo{volume}{391}}, \bibinfo{pages}{176} (\bibinfo{year}{1997}).

\bibitem[{\citenamefont{Giessel et~al.}(1998)\citenamefont{Giessel, Schaff,
  Hirschmugl, Fernandez, Schindler, Theobald, Bao, Lindsay, Berndt, Bradshaw, Baddeley, Lee, Lambert and Woodruff }}]{ss:Giessel:406}
\bibinfo{author}{\bibfnamefont{T.}~\bibnamefont{Giessel}},
  \bibinfo{author}{\bibfnamefont{O.}~\bibnamefont{Schaff}},
  \bibinfo{author}{\bibfnamefont{C.~J.} \bibnamefont{Hirschmugl}},
  \bibinfo{author}{\bibfnamefont{V.}~\bibnamefont{Fernandez}},
  \bibinfo{author}{\bibfnamefont{K.~M.} \bibnamefont{Schindler}},
 \bibinfo{author}{\bibfnamefont{A.}~\bibnamefont{Theobald}},
  \bibinfo{author}{\bibfnamefont{S.}~\bibnamefont{Bao}},
  \bibinfo{author}{\bibfnamefont{R.}~\bibnamefont{Lindsay}},
  \bibinfo{author}{\bibfnamefont{W.}~\bibnamefont{Berndt}},
  \bibinfo{author}{\bibfnamefont{A.~M.} \bibnamefont{Bradshaw}},
   \bibinfo{author}{\bibfnamefont{C.}~\bibnamefont{Baddeley}},
 \bibinfo{author}{\bibfnamefont{A.~F.} \bibnamefont{Lee}},
 \bibinfo{author}{\bibfnamefont{R.~M.} \bibnamefont{Lambert}} and
 \bibinfo{author}{\bibfnamefont{D.~P.} \bibnamefont{Woodruff}}, \bibinfo{journal}{Surf. Sci.}
  \textbf{\bibinfo{volume}{406}}, \bibinfo{pages}{90} (\bibinfo{year}{1998}).

\bibitem[{\citenamefont{Ohtani et~al.}(1987)\citenamefont{Ohtani, {Van }Hove,
  and Somorjai}}]{ss:Ohtani:187}
\bibinfo{author}{\bibfnamefont{H.}~\bibnamefont{Ohtani}},
  \bibinfo{author}{\bibfnamefont{M.~A.} \bibnamefont{{Van }Hove}},
  \bibnamefont{and} \bibinfo{author}{\bibfnamefont{G.~A.}
  \bibnamefont{Somorjai}}, \bibinfo{journal}{Surf. Sci.}
  \textbf{\bibinfo{volume}{187}}, \bibinfo{pages}{372} (\bibinfo{year}{1987}).

\bibitem[{\citenamefont{Blackman et~al.}(1988)\citenamefont{Blackman, Xu,
  Ogletree, {Van }Hove, and Somorjai}}]{ss:Blackman:61}
\bibinfo{author}{\bibfnamefont{G.~S.} \bibnamefont{Blackman}},
  \bibinfo{author}{\bibfnamefont{M.~L.} \bibnamefont{Xu}},
  \bibinfo{author}{\bibfnamefont{D.~F.} \bibnamefont{Ogletree}},
  \bibinfo{author}{\bibfnamefont{M.~A.} \bibnamefont{{Van }Hove}},
  \bibnamefont{and} \bibinfo{author}{\bibfnamefont{G.~A.}
  \bibnamefont{Somorjai}}, \bibinfo{journal}{Surf. Sci.}
  \textbf{\bibinfo{volume}{61}}, \bibinfo{pages}{2353} (\bibinfo{year}{1988}).

\bibitem[{\citenamefont{Beitel et~al.}(1996)\citenamefont{Beitel, Laskov,
  Oosterbeek, and Kuipers}}]{jpc:Beitel:100}
\bibinfo{author}{\bibfnamefont{G.~A.} \bibnamefont{Beitel}},
  \bibinfo{author}{\bibfnamefont{A.}~\bibnamefont{Laskov}},
  \bibinfo{author}{\bibfnamefont{H.}~\bibnamefont{Oosterbeek}},
  \bibnamefont{and} \bibinfo{author}{\bibfnamefont{W.~W.}
  \bibnamefont{Kuipers}}, \bibinfo{journal}{J. Phys. Chem.}
  \textbf{\bibinfo{volume}{100}}, \bibinfo{pages}{12494}
  (\bibinfo{year}{1996}).

\bibitem[{\citenamefont{Hirschmugl et~al.}(1990)\citenamefont{Hirschmugl,
  Williams, Hoffmann, and Chabal}}]{jesrp:Hirschmugl:54}
\bibinfo{author}{\bibfnamefont{C.~J.} \bibnamefont{Hirschmugl}},
  \bibinfo{author}{\bibfnamefont{G.~P.} \bibnamefont{Williams}},
  \bibinfo{author}{\bibfnamefont{F.~M.} \bibnamefont{Hoffmann}},
  \bibnamefont{and} \bibinfo{author}{\bibfnamefont{Y.~J.}
  \bibnamefont{Chabal}}, \bibinfo{journal}{J. Electron Spectrosc. Relat.
  Phenom.} \textbf{\bibinfo{volume}{54/55}}, \bibinfo{pages}{109}
  (\bibinfo{year}{1990}).

\bibitem[{\citenamefont{Raval et~al.}(1988)\citenamefont{Raval, Parker, Pemble,
  Hollins, Pritchard, and Chester}}]{ss:Raval:203}
\bibinfo{author}{\bibfnamefont{R.}~\bibnamefont{Raval}},
  \bibinfo{author}{\bibfnamefont{S.~F.} \bibnamefont{Parker}},
  \bibinfo{author}{\bibfnamefont{M.~E.} \bibnamefont{Pemble}},
  \bibinfo{author}{\bibfnamefont{P.}~\bibnamefont{Hollins}},
  \bibinfo{author}{\bibfnamefont{J.}~\bibnamefont{Pritchard}},
  \bibnamefont{and} \bibinfo{author}{\bibfnamefont{M.~A.}
  \bibnamefont{Chester}}, \bibinfo{journal}{Surf. Sci.}
  \textbf{\bibinfo{volume}{203}}, \bibinfo{pages}{353} (\bibinfo{year}{1988}).

\bibitem[{\citenamefont{Eve and McCash}(1999)}]{cpl:Eve:313}
\bibinfo{author}{\bibfnamefont{J.~K.} \bibnamefont{Eve}} \bibnamefont{and}
  \bibinfo{author}{\bibfnamefont{E.~M.} \bibnamefont{McCash}},
  \bibinfo{journal}{Chem. Phys. Lett.} \textbf{\bibinfo{volume}{313}},
  \bibinfo{pages}{575} (\bibinfo{year}{1999}).

\bibitem[{\citenamefont{Thomas and Weinberg}(1979)}]{jcp:Thomas:70}
\bibinfo{author}{\bibfnamefont{G.~E.} \bibnamefont{Thomas}} \bibnamefont{and}
  \bibinfo{author}{\bibfnamefont{W.~H.} \bibnamefont{Weinberg}},
  \bibinfo{journal}{J. Chem. Phys.} \textbf{\bibinfo{volume}{70}},
  \bibinfo{pages}{954} (\bibinfo{year}{1979}).

\bibitem[{\citenamefont{Jacob and Person}(1997)}]{prl:Jacob:78}
\bibinfo{author}{\bibfnamefont{P.}~\bibnamefont{Jacob}} \bibnamefont{and}
  \bibinfo{author}{\bibfnamefont{B.~N.~J.} \bibnamefont{Person}},
  \bibinfo{journal}{Phys. Rev. Lett.} \textbf{\bibinfo{volume}{78}},
  \bibinfo{pages}{3503} (\bibinfo{year}{1997}).

\bibitem[{\citenamefont{He et~al.}(1996)\citenamefont{He, Dietrich, and
  Jacobi}}]{ss:He:345}
\bibinfo{author}{\bibfnamefont{P.}~\bibnamefont{He}},
  \bibinfo{author}{\bibfnamefont{H.}~\bibnamefont{Dietrich}}, \bibnamefont{and}
  \bibinfo{author}{\bibfnamefont{K.}~\bibnamefont{Jacobi}},
  \bibinfo{journal}{Surf. Sci.} \textbf{\bibinfo{volume}{345}},
  \bibinfo{pages}{241} (\bibinfo{year}{1996}).

\bibitem[{\citenamefont{Dubois and Somorjai}(1979)}]{ss:Dubois:91}
\bibinfo{author}{\bibfnamefont{L.~H.} \bibnamefont{Dubois}} \bibnamefont{and}
  \bibinfo{author}{\bibfnamefont{G.~A.} \bibnamefont{Somorjai}},
  \bibinfo{journal}{Surf. Sci.} \textbf{\bibinfo{volume}{91}},
  \bibinfo{pages}{514} (\bibinfo{year}{1979}).

\bibitem[{\citenamefont{Smedh et~al.}(2001)\citenamefont{Smedh, Beutler,
  Ramsvik, Nyholm, Borg, Andersen, Duschek, Sock, Netzer, and
  Ramsey}}]{ss:Smedh:491:99}
\bibinfo{author}{\bibfnamefont{M.}~\bibnamefont{Smedh}},
  \bibinfo{author}{\bibfnamefont{A.}~\bibnamefont{Beutler}},
  \bibinfo{author}{\bibfnamefont{T.}~\bibnamefont{Ramsvik}},
  \bibinfo{author}{\bibfnamefont{R.}~\bibnamefont{Nyholm}},
  \bibinfo{author}{\bibfnamefont{M.}~\bibnamefont{Borg}},
  \bibinfo{author}{\bibfnamefont{J.~N.} \bibnamefont{Andersen}},
  \bibinfo{author}{\bibfnamefont{R.}~\bibnamefont{Duschek}},
  \bibinfo{author}{\bibfnamefont{M.}~\bibnamefont{Sock}},
  \bibinfo{author}{\bibfnamefont{F.~P.} \bibnamefont{Netzer}},
  \bibnamefont{and} \bibinfo{author}{\bibfnamefont{M.~G.}
  \bibnamefont{Ramsey}}, \bibinfo{journal}{Surf. Sci.}
  \textbf{\bibinfo{volume}{491}}, \bibinfo{pages}{99} (\bibinfo{year}{2001}).

\bibitem[{\citenamefont{Bradshaw and Hoffmann}(1978)}]{ss:Bradshaw:72}
\bibinfo{author}{\bibfnamefont{A.~M.} \bibnamefont{Bradshaw}} \bibnamefont{and}
  \bibinfo{author}{\bibfnamefont{F.~M.} \bibnamefont{Hoffmann}},
  \bibinfo{journal}{Surf. Sci.} \textbf{\bibinfo{volume}{72}},
  \bibinfo{pages}{513} (\bibinfo{year}{1978}).

\bibitem[{\citenamefont{Surnev et~al.}(2000)\citenamefont{Surnev, Sock, Ramsey,
  Netzer, Wiklund, Borg, and Andersen}}]{ss:Surnev:470}
\bibinfo{author}{\bibfnamefont{S.}~\bibnamefont{Surnev}},
  \bibinfo{author}{\bibfnamefont{M.}~\bibnamefont{Sock}},
  \bibinfo{author}{\bibfnamefont{M.~G.} \bibnamefont{Ramsey}},
  \bibinfo{author}{\bibfnamefont{F.~P.} \bibnamefont{Netzer}},
  \bibinfo{author}{\bibfnamefont{M.}~\bibnamefont{Wiklund}},
  \bibinfo{author}{\bibfnamefont{M.}~\bibnamefont{Borg}}, \bibnamefont{and}
  \bibinfo{author}{\bibfnamefont{J.~N.} \bibnamefont{Andersen}},
  \bibinfo{journal}{Surf. Sci.} \textbf{\bibinfo{volume}{470}},
  \bibinfo{pages}{171} (\bibinfo{year}{2000}).

\bibitem[{\citenamefont{Hansen et~al.}(1991)\citenamefont{Hansen, Bertolo, and
  Jacobi}}]{ss:Hansen:253}
\bibinfo{author}{\bibfnamefont{W.}~\bibnamefont{Hansen}},
  \bibinfo{author}{\bibfnamefont{M.}~\bibnamefont{Bertolo}}, \bibnamefont{and}
  \bibinfo{author}{\bibfnamefont{K.}~\bibnamefont{Jacobi}},
  \bibinfo{journal}{Surf. Sci.} \textbf{\bibinfo{volume}{253}},
  \bibinfo{pages}{1} (\bibinfo{year}{1991}).

\bibitem[{\citenamefont{Schick et~al.}(1996)\citenamefont{Schick, Lauterbach,
  and Weinberg}}]{jvst:Schick:14}
\bibinfo{author}{\bibfnamefont{M.}~\bibnamefont{Schick}},
  \bibinfo{author}{\bibfnamefont{J.}~\bibnamefont{Lauterbach}},
  \bibnamefont{and} \bibinfo{author}{\bibfnamefont{W.~H.}
  \bibnamefont{Weinberg}}, \bibinfo{journal}{J. Vac. Sci. Technol. A}
  \textbf{\bibinfo{volume}{14}}, \bibinfo{pages}{1448} (\bibinfo{year}{1996}).

\bibitem[{\citenamefont{Lauterbach et~al.}(1996)\citenamefont{Lauterbach,
  Boyle, Schick, Mitchell, Meng, and Weinberg}}]{ss:Lauterbach:350}
\bibinfo{author}{\bibfnamefont{J.}~\bibnamefont{Lauterbach}},
  \bibinfo{author}{\bibfnamefont{R.~W.} \bibnamefont{Boyle}},
  \bibinfo{author}{\bibfnamefont{M.}~\bibnamefont{Schick}},
  \bibinfo{author}{\bibfnamefont{W.~J.} \bibnamefont{Mitchell}},
  \bibinfo{author}{\bibfnamefont{B.}~\bibnamefont{Meng}}, \bibnamefont{and}
  \bibinfo{author}{\bibfnamefont{W.~H.} \bibnamefont{Weinberg}},
  \bibinfo{journal}{Surf. Sci.} \textbf{\bibinfo{volume}{350}},
  \bibinfo{pages}{32} (\bibinfo{year}{1996}).

\bibitem[{\citenamefont{Kung et~al.}(2000)\citenamefont{Kung, Chen, Wei, Shen,
  and Somorjai}}]{ss:Kung:463}
\bibinfo{author}{\bibfnamefont{K.~Y.} \bibnamefont{Kung}},
  \bibinfo{author}{\bibfnamefont{P.}~\bibnamefont{Chen}},
  \bibinfo{author}{\bibfnamefont{F.}~\bibnamefont{Wei}},
  \bibinfo{author}{\bibfnamefont{Y.~R.} \bibnamefont{Shen}}, \bibnamefont{and}
  \bibinfo{author}{\bibfnamefont{G.~A.} \bibnamefont{Somorjai}},
  \bibinfo{journal}{Surf. Sci.} \textbf{\bibinfo{volume}{463}},
  \bibinfo{pages}{L627} (\bibinfo{year}{2000}).

\bibitem[{\citenamefont{Surman et~al.}(2002)\citenamefont{Surman, Hagans,
  Wilson, Baily, and Russell}}]{ss:Surman:511}
\bibinfo{author}{\bibfnamefont{M.}~\bibnamefont{Surman}},
  \bibinfo{author}{\bibfnamefont{P.~L.} \bibnamefont{Hagans}},
  \bibinfo{author}{\bibfnamefont{N.~E.} \bibnamefont{Wilson}},
  \bibinfo{author}{\bibfnamefont{C.~J.} \bibnamefont{Baily}}, \bibnamefont{and}
  \bibinfo{author}{\bibfnamefont{A.~E.} \bibnamefont{Russell}},
  \bibinfo{journal}{Surf. Sci.} \textbf{\bibinfo{volume}{511}},
  \bibinfo{pages}{L303} (\bibinfo{year}{2002}).

\bibitem[{\citenamefont{Schweizer et~al.}(1989)\citenamefont{Schweizer,
  Persson, T{\"u}shaus, Hoge, and Bradshaw}}]{ss:Schweizer:213}
\bibinfo{author}{\bibfnamefont{E.}~\bibnamefont{Schweizer}},
  \bibinfo{author}{\bibfnamefont{B.~N.~J.} \bibnamefont{Persson}},
  \bibinfo{author}{\bibfnamefont{M.}~\bibnamefont{T{\"u}shaus}},
  \bibinfo{author}{\bibfnamefont{D.}~\bibnamefont{Hoge}}, \bibnamefont{and}
  \bibinfo{author}{\bibfnamefont{A.~M.} \bibnamefont{Bradshaw}},
  \bibinfo{journal}{Surf. Sci.} \textbf{\bibinfo{volume}{213}},
  \bibinfo{pages}{49} (\bibinfo{year}{1989}).

\bibitem[{\citenamefont{Nekrylova and Harrison}(1996)}]{cp:Nekrylova:205}
\bibinfo{author}{\bibfnamefont{J.~V.} \bibnamefont{Nekrylova}}
  \bibnamefont{and} \bibinfo{author}{\bibfnamefont{I.}~\bibnamefont{Harrison}},
  \bibinfo{journal}{Chem. Phys.} \textbf{\bibinfo{volume}{205}},
  \bibinfo{pages}{37} (\bibinfo{year}{1996}).

\bibitem[{\citenamefont{Yoshinobu and Kawai}(1996)}]{ss:Yoshinobu:363}
\bibinfo{author}{\bibfnamefont{J.}~\bibnamefont{Yoshinobu}} \bibnamefont{and}
  \bibinfo{author}{\bibfnamefont{M.}~\bibnamefont{Kawai}},
  \bibinfo{journal}{Surf. Sci.} \textbf{\bibinfo{volume}{363}},
  \bibinfo{pages}{105} (\bibinfo{year}{1996}).

\bibitem[{\citenamefont{Heyden and Bradshaw}(1983)}]{ss:Heyden:125}
\bibinfo{author}{\bibfnamefont{B.~E.} \bibnamefont{Heyden}} \bibnamefont{and}
  \bibinfo{author}{\bibfnamefont{A.~M.} \bibnamefont{Bradshaw}},
  \bibinfo{journal}{Surf. Sci.} \textbf{\bibinfo{volume}{125}},
  \bibinfo{pages}{787} (\bibinfo{year}{1983}).

\bibitem[{\citenamefont{Nekrylova et~al.}(1993)\citenamefont{Nekrylova, French,
  Artsyukhovich, Ukraintsev, and Harrison}}]{ssl:Nekrylova:295}
\bibinfo{author}{\bibfnamefont{J.~V.} \bibnamefont{Nekrylova}},
  \bibinfo{author}{\bibfnamefont{C.}~\bibnamefont{French}},
  \bibinfo{author}{\bibfnamefont{A.~N.} \bibnamefont{Artsyukhovich}},
  \bibinfo{author}{\bibfnamefont{V.~A.} \bibnamefont{Ukraintsev}},
  \bibnamefont{and} \bibinfo{author}{\bibfnamefont{I.}~\bibnamefont{Harrison}},
  \bibinfo{journal}{Sur. Sci. Lett.} \textbf{\bibinfo{volume}{295}},
  \bibinfo{pages}{L987} (\bibinfo{year}{1993}).

\bibitem[{\citenamefont{F\"ohlisch et~al.}(2000)\citenamefont{F\"ohlisch,
  Nyberg, Hasselstr\"om, Karis, Pettersson, and Nilsson}}]{prl:Foehlisch:85}
\bibinfo{author}{\bibfnamefont{A.}~\bibnamefont{F\"ohlisch}},
  \bibinfo{author}{\bibfnamefont{M.}~\bibnamefont{Nyberg}},
  \bibinfo{author}{\bibfnamefont{J.}~\bibnamefont{Hasselstr\"om}},
  \bibinfo{author}{\bibfnamefont{O.}~\bibnamefont{Karis}},
  \bibinfo{author}{\bibfnamefont{L.~G.~M.} \bibnamefont{Pettersson}},
  \bibnamefont{and} \bibinfo{author}{\bibfnamefont{A.}~\bibnamefont{Nilsson}},
  \bibinfo{journal}{Phys. Rev. Lett.} \textbf{\bibinfo{volume}{85}},
  \bibinfo{pages}{3309} (\bibinfo{year}{2000}).

\bibitem[{\citenamefont{Kresse et~al.}(2003)\citenamefont{Kresse, Gil, and
  Sautet}}]{prb:Kresse:68}
\bibinfo{author}{\bibfnamefont{G.}~\bibnamefont{Kresse}},
  \bibinfo{author}{\bibfnamefont{A.}~\bibnamefont{Gil}}, \bibnamefont{and}
  \bibinfo{author}{\bibfnamefont{P.}~\bibnamefont{Sautet}},
  \bibinfo{journal}{Phys. Rev. B} \textbf{\bibinfo{volume}{68}},
  \bibinfo{pages}{073401} (\bibinfo{year}{2003}).

\bibitem[{\citenamefont{Bagus and Pacchioni}(1972)}]{ss:Bagus:278}
\bibinfo{author}{\bibfnamefont{P.~S.} \bibnamefont{Bagus}} \bibnamefont{and}
  \bibinfo{author}{\bibfnamefont{G.}~\bibnamefont{Pacchioni}},
  \bibinfo{journal}{Surf. Sci.} \textbf{\bibinfo{volume}{278}},
  \bibinfo{pages}{427} (\bibinfo{year}{1972}).

\bibitem[{\citenamefont{Hoffmann}(1988)}]{rmp:Hoffmann:60}
\bibinfo{author}{\bibfnamefont{R.}~\bibnamefont{Hoffmann}},
  \bibinfo{journal}{Rev. Mod. Phys.} \textbf{\bibinfo{volume}{60}},
  \bibinfo{pages}{601} (\bibinfo{year}{1988}).

\bibitem[{\citenamefont{Illas et~al.}(1997)\citenamefont{Illas, Zurita, Rubio,
  and M\'{a}rquez}}]{ss:Illas:376}
\bibinfo{author}{\bibfnamefont{F.}~\bibnamefont{Illas}},
  \bibinfo{author}{\bibfnamefont{S.}~\bibnamefont{Zurita}},
  \bibinfo{author}{\bibfnamefont{J.}~\bibnamefont{Rubio}}, \bibnamefont{and}
  \bibinfo{author}{\bibfnamefont{A.~M.} \bibnamefont{M\'{a}rquez}},
  \bibinfo{journal}{Surf. Sci.} \textbf{\bibinfo{volume}{376}},
  \bibinfo{pages}{279} (\bibinfo{year}{1997}).

\bibitem[{\citenamefont{Aizawa and Tsuneyuki}(1998)}]{ss:Aizawa:399}
\bibinfo{author}{\bibfnamefont{H.}~\bibnamefont{Aizawa}} \bibnamefont{and}
  \bibinfo{author}{\bibfnamefont{S.}~\bibnamefont{Tsuneyuki}},
  \bibinfo{journal}{Surf. Sci.} \textbf{\bibinfo{volume}{399}},
  \bibinfo{pages}{L364} (\bibinfo{year}{1998}).

\bibitem[{\citenamefont{Gumhalter et~al.}(1988)\citenamefont{Gumhalter,
  Wandelt, and Avouris}}]{prb:Gumhalter:37}
\bibinfo{author}{\bibfnamefont{B.}~\bibnamefont{Gumhalter}},
  \bibinfo{author}{\bibfnamefont{K.}~\bibnamefont{Wandelt}}, \bibnamefont{and}
  \bibinfo{author}{\bibfnamefont{P.}~\bibnamefont{Avouris}},
  \bibinfo{journal}{Phys. Rev. B} \textbf{\bibinfo{volume}{37}},
  \bibinfo{pages}{8048} (\bibinfo{year}{1988}).

\bibitem[{\citenamefont{Ohnishi and Watari}(1994)}]{prb:Ohnishi:49}
\bibinfo{author}{\bibfnamefont{S.}~\bibnamefont{Ohnishi}} \bibnamefont{and}
  \bibinfo{author}{\bibfnamefont{N.}~\bibnamefont{Watari}},
  \bibinfo{journal}{Phys. Rev. B} \textbf{\bibinfo{volume}{49}},
  \bibinfo{pages}{14619} (\bibinfo{year}{1994}).

\bibitem[{\citenamefont{Nilsson et~al.}(1997)\citenamefont{Nilsson, Weinelt,
  Wiell, Bennich, Karis, and Wassdahl}}]{prl:Nilsson:78}
\bibinfo{author}{\bibfnamefont{A.}~\bibnamefont{Nilsson}},
  \bibinfo{author}{\bibfnamefont{M.}~\bibnamefont{Weinelt}},
  \bibinfo{author}{\bibfnamefont{T.}~\bibnamefont{Wiell}},
  \bibinfo{author}{\bibfnamefont{P.}~\bibnamefont{Bennich}},
  \bibinfo{author}{\bibfnamefont{O.}~\bibnamefont{Karis}}, \bibnamefont{and}
  \bibinfo{author}{\bibfnamefont{N.}~\bibnamefont{Wassdahl}},
  \bibinfo{journal}{Phys. Rev. Lett.} \textbf{\bibinfo{volume}{78}},
  \bibinfo{pages}{2847} (\bibinfo{year}{1997}).

\bibitem[{\citenamefont{Hu et~al.}(1995)\citenamefont{Hu, King, Lee, and
  Payne}}]{cpl:Hu:246}
\bibinfo{author}{\bibfnamefont{P.}~\bibnamefont{Hu}},
  \bibinfo{author}{\bibfnamefont{D.~A.} \bibnamefont{King}},
  \bibinfo{author}{\bibfnamefont{M.-H.} \bibnamefont{Lee}}, \bibnamefont{and}
  \bibinfo{author}{\bibfnamefont{M.~C.} \bibnamefont{Payne}},
  \bibinfo{journal}{Chem. Phys. Lett.} \textbf{\bibinfo{volume}{246}},
  \bibinfo{pages}{73} (\bibinfo{year}{1995}).

\bibitem[{\citenamefont{Lu et~al.}(2002)\citenamefont{Lu, Lee, Masel,
  Wieckowski, and Rice}}]{jpca:Lu:106}
\bibinfo{author}{\bibfnamefont{C.}~\bibnamefont{Lu}},
  \bibinfo{author}{\bibfnamefont{I.~C.} \bibnamefont{Lee}},
  \bibinfo{author}{\bibfnamefont{R.~I.} \bibnamefont{Masel}},
  \bibinfo{author}{\bibfnamefont{A.}~\bibnamefont{Wieckowski}},
  \bibnamefont{and} \bibinfo{author}{\bibfnamefont{C.}~\bibnamefont{Rice}},
  \bibinfo{journal}{J. Phys. Chem. A} \textbf{\bibinfo{volume}{106}},
  \bibinfo{pages}{3084} (\bibinfo{year}{2002}).

\bibitem[{\citenamefont{Geschke et~al.}(2001)\citenamefont{Geschke, Ba{\c
  s}tu{\v g}, Jacob, Fritzsche, Sepp, Fricke, Varga, and
  Anton}}]{prb:Geschke:64}
\bibinfo{author}{\bibfnamefont{D.}~\bibnamefont{Geschke}},
  \bibinfo{author}{\bibfnamefont{T.}~\bibnamefont{Ba{\c s}tu{\v g}}},
  \bibinfo{author}{\bibfnamefont{T.}~\bibnamefont{Jacob}},
  \bibinfo{author}{\bibfnamefont{S.}~\bibnamefont{Fritzsche}},
  \bibinfo{author}{\bibfnamefont{W.-D.} \bibnamefont{Sepp}},
  \bibinfo{author}{\bibfnamefont{B.}~\bibnamefont{Fricke}},
  \bibinfo{author}{\bibfnamefont{S.}~\bibnamefont{Varga}}, \bibnamefont{and}
  \bibinfo{author}{\bibfnamefont{J.}~\bibnamefont{Anton}},
  \bibinfo{journal}{Phys. Rev. B} \textbf{\bibinfo{volume}{64}},
  \bibinfo{pages}{235411} (\bibinfo{year}{2001}).

\bibitem[{\citenamefont{Grinberg et~al.}(2002)\citenamefont{Grinberg,
  Yourdshahyan, and Rappe}}]{jcp:Grinberg:117}
\bibinfo{author}{\bibfnamefont{I.}~\bibnamefont{Grinberg}},
  \bibinfo{author}{\bibfnamefont{Y.}~\bibnamefont{Yourdshahyan}},
  \bibnamefont{and} \bibinfo{author}{\bibfnamefont{A.~M.} \bibnamefont{Rappe}},
  \bibinfo{journal}{J. Chem. Phys.} \textbf{\bibinfo{volume}{117}},
  \bibinfo{pages}{2264} (\bibinfo{year}{2002}).

\bibitem[{\citenamefont{Olsen et~al.}(2003)\citenamefont{Olsen, Philipsen, and
  Baerends}}]{jcp:Olsen:119}
\bibinfo{author}{\bibfnamefont{R.~A.} \bibnamefont{Olsen}},
  \bibinfo{author}{\bibfnamefont{P.~H.~T.} \bibnamefont{Philipsen}},
  \bibnamefont{and} \bibinfo{author}{\bibfnamefont{E.~J.}
  \bibnamefont{Baerends}}, \bibinfo{journal}{J. Chem. Phys.}
  \textbf{\bibinfo{volume}{119}}, \bibinfo{pages}{4522} (\bibinfo{year}{2003}).

\bibitem[{\citenamefont{Gil et~al.}(2003)\citenamefont{Gil, Clotet, Ricart,
  Kresse, Garc{\'i}a{-}Hern{\'a}ndez, R{\"o}sch, and Sautet}}]{ss:Gil:530}
\bibinfo{author}{\bibfnamefont{A.}~\bibnamefont{Gil}},
  \bibinfo{author}{\bibfnamefont{A.}~\bibnamefont{Clotet}},
  \bibinfo{author}{\bibfnamefont{J.~M.} \bibnamefont{Ricart}},
  \bibinfo{author}{\bibfnamefont{G.}~\bibnamefont{Kresse}},
  \bibinfo{author}{\bibfnamefont{M.}~\bibnamefont{Garc{\'i}a{-}Hern{\'a}ndez}},
  \bibinfo{author}{\bibfnamefont{N.}~\bibnamefont{R{\"o}sch}},
  \bibnamefont{and} \bibinfo{author}{\bibfnamefont{P.}~\bibnamefont{Sautet}},
  \bibinfo{journal}{Surf. Sci.} \textbf{\bibinfo{volume}{530}},
  \bibinfo{pages}{71} (\bibinfo{year}{2003}).

\end{thebibliography}

\end{document}